\documentclass[aps,prd,twocolumn,groupedaddress,floatfix,nofootinbib]{revtex4-1}

\usepackage[utf8]{inputenc}
\usepackage[T1]{fontenc}
\usepackage{graphicx}
\usepackage{color}
\usepackage{amsmath}
\usepackage{amssymb}
\usepackage{hyperref}

\begin{document}

\title{Thermal fluctuations on the freeze-out surface of heavy-ion collisions\\and their impact on particle correlations}

\author{Adrian Skasberg Aasen}
\email[]{aasen@kip.uni-heidelberg.de}

\affiliation{Kirchhoff-Institut f\"ur Physik, Universit\"at Heidelberg,
Im Neuenheimer Feld 227, 69120 Heidelberg, Germany}

\author{Stefan Floerchinger}
\email[]{stefan.floerchinger@uni-jena.de}
\affiliation{Theoretisch-Physikalisches Institut, Friedrich-Schiller-Universität Jena, Max-Wien-Platz 1, 07743 Jena, Germany}

\author{Giuliano Giacalone}
\email[]{giacalone@thphys.uni-heidelberg.de}

\affiliation{Institut f\"{u}r Theoretische Physik, Universit\"{a}t Heidelberg, Philosophenweg 16, 69120 Heidelberg, Germany}

\author{Deniz Guenduez}
\email[]{guenduez@mpi-hd.mpg.de}
\affiliation{Max Planck Institute for Nuclear Physics,  Saupfercheckweg 1, 69117 Heidelberg, Germany}

\begin{abstract}
Particle momentum distributions originating from a quark-gluon plasma as produced in high-energy nuclear collisions can be influenced by thermal fluctuations in fluid dynamic fields. We study this effect by generalising the commonly used kinetic freeze-out prescription by allowing for small fluctuations around an average in fluid velocity, chemical potentials and temperature. This leads to the appearance of specific two-body momentum correlations. Combining a blast-wave parametrization of the kinetic freeze-out surface with the thermal correlation functions of an ideal resonance gas, we perform an exploratory study of angular net-charge correlations induced by thermal fluctuations around vanishing chemical potential. We note a diffusion of the near-side peak around $\Delta y=\Delta\phi=0$ induced by variances of different chemical potentials, which could be investigated experimentally.
\end{abstract}

\maketitle

\section{Introduction\label{sec:Introduction}}

The phenomenology of ultrarelativistic heavy-ion collisions is centered around the understanding of the  evolution of the quark-gluon plasma (QGP) that emerges in the interaction of two nuclei at high energy \cite{Braun-Munzinger:2007edi,Busza:2018rrf}. Two decades of experiments have established an effective description of the QGP based on relativistic fluid dynamics \cite{Romatschke:2017ejr}. The system reaches local thermal equilibrium over a time scale of order 1 fm/$c$ \cite{Schlichting:2019abc,Berges:2020fwq}, and it expands as a relativistic fluid for about $10$ fm/$c$ before decoupling to hadrons that eventually freely stream to the detectors. 

Fluctuations on top of this picture play a major role in the phenomenology of heavy-ion collisions. They are mainly of two kinds. There are initial-state fluctuations \cite{Luzum:2013yya}, originating from the fact that the structure of the colliding ions and nucleons fluctuates on an event-by-event basis. They yield, notably, a landscape of peaks and valleys in the created QGP energy density, with nontrivial implications for the subsequent fluid evolution and its signatures. There are, in addition, final-state fluctuations arising from the process of particlization of the fluid, driven by the fact that the QGP decouples to a finite number of hadrons. Of particular interest are the fluctuations of conserved quantities (e.g. baryon number and electric charge), which fluctuate locally, as well as globally within the finite phase-space coverage of the detectors. They might be sensitive to the details of the QGP-hadron transition and potentially to associated critical phenomena \cite{Cheng:2008zh,Fukushima:2010bq,Bzdak:2019pkr,Bluhm:2020mpc}.

Besides initial- and final-state fluctuations, thermal (or hydrodynamic) fluctuations engendered by the expansion dynamics are also present, in agreement with the dissipation-fluctuation theorem and the fact that the QGP is not a perfect fluid. These fluctuations are the focus of this manuscript. Hydrodynamic fluctuations have been studied in the context of QGP expansion since a long time ago \cite{Kapusta:2011gt,Young:2013fka,Ling:2013ksb,Young:2014pka,Murase:2015oie,Yan:2015lfa,Nagai:2016wyx,Gavin:2016hmv,Akamatsu:2016llw,Sakai:2017rfi,Kapusta:2017hfi,Chattopadhyay:2017rgh,Stephanov:2017ghc,Singh:2018dpk,Akamatsu:2018vjr,An:2019osr,Murase:2019cwc,An:2019csj,De:2020yyx,Sakai:2020pjw,Chao:2020kcf,Sakai:2021pev}. Their implementation goes typically via the addition of a noise term in the relativistic hydrodynamic equations, and their phenomenological consequences have also been investigated quantitatively. 

In this paper, we study how correlation functions of hadrons emerging from the final stages of the QGP are impacted by thermal fluctuations in the fluid fields that describe the system at the intersection between a fluid and a particle description, i.e., on the so-called \textit{freeze-out} surface. We ask what two-body correlations are engendered in momentum space if one relaxes the common description of the freeze-out surface with unique fluid fields and fixed freeze-out temperature, but takes thermal fluctuations in fluid velocity, chemical potentials and temperature into account. 

Our goal is in a sense simpler with respect to previous works, as we shall not aim at evolving thermal fluctuations dynamically over the fluid evolution. We take the standard fluid evolution equations to describe expectation values and to be renormalized in that sense, already.  However, our study will permit us to point out generic phenomenological signatures of thermal fluctuations that have not been discussed yet in the literature, and to do so in a simple analytical framework. Our analysis focuses, in particular, on the emergence of differential net-charge fluctuations, depending on transverse momentum and rapidity cuts, which may be straightforwardly measured in experiments. 

This manuscript is organized as follows. In Sec.~\ref{sec:2} we briefly recall the physical picture of kinetic freeze-out in heavy-ion collisions, and the formalism adopted to describe it. In Sec.~\ref{sec:3}, we present our main result, namely, the derivation of a relation between two-particle correlations in momentum space arising from the QGP particlization and two-point correlation functions in the underlying fluid fields, which we achieve by allowing for small fluctuations around the average freeze-out temperature. We discuss, hence, the impact of these thermal fluctuations on fully-differential momentum-space observables typically investigated in heavy-ion collision experiments. We move on, then, to an application. In Sec.~\ref{sec:4}, we construct a model for the fluctuating kinetic freeze-out surface of heavy-ion collisions. The average hypersurface and the average yields in momentum space are modeled via a blast-wave parametrization, while thermal fluctuations on the freeze-out surface are implemented from the two-point thermal correlation functions of an ideal resonance gas. In Sec.~\ref{sec:5}, we use this model to perform a numerical study of the imprints of thermal fluctuations at freeze-out. We compute the angular correlations of net electric charge and net baryon number,  emphasizing  the nontrivial role played by the variances of chemical potentials. Section~\ref{sec:6} is left for conclusions and an outlook on possible applications of our results.

\section{\label{sec:2}  kinetic freeze-out surface in heavy-ion collisions}

We define the kinetic freeze-out surface as the space-time hypersurface at which particle scattering ceases to be effective in maintaining local kinetic equilibrium. It is sometimes referred  to as the ``surface of last scattering''. As local thermal equilibrium is lost, application of fluid dynamics is no longer justified, and a description in terms of particles and their distribution functions sets in.

\subsection{\label{sec:The Cooper-Frye prescription}The Cooper-Frye prescription}

To make this transition, one can view the freeze-out surface as an extended source of particles with local occupation number in phase space
\begin{equation}
    \frac{dN_i}{d^3pd^3x}=\frac{1}{(2\pi)^3}f_i(x,p).
\end{equation}
Here $f_i$ is the particle distribution function for particle species $i$, $x=(t,\mathbf{x})$ denotes the space-time coordinate on the freeze-out surface, while $p=(E,\mathbf{p})$ is the particle momentum. The Lorentz invariant single-particle distribution in momentum space, $E dN_{i}/d^{3}p$, can be obtained for a given shape of the freeze-out surface and the distribution function $f_i(x, p)$ from the Cooper-Frye integral \cite{Cooper:1974mv},
\begin{equation}
    \label{eq:CF1P}
    E\frac{dN_{i}}{d^{3}p}=-\frac{1}{(2\pi)^{3}}\int \limits_{\Sigma_{f}} d\Sigma_{\mu} \,  p^{\mu} f_{i}(x,p).
\end{equation}
Here $d\Sigma_\mu$ is the freeze-out surface (dubbed $\Sigma_f$) element and the additional minus sign is due to our metric signature $(-,+,+,+)$ and the orientation of the hypersurface element to the future. 

In the regime where the fluid description intersects with a particle description, the distribution function depends on the momentum $p^\mu$ of the final state particles and the position on the freeze-out surface through the underlying fluid fields,
\begin{equation}
    f_{i}=f_{i}(p^{\mu}, u^{\mu}(x), T(x),\mu_{j}(x), \pi^{\mu \nu}(x),\pi_{\mathrm{bulk}}(x)),
\label{eq:distributionFunction}
\end{equation}
where we have introduced the fluid velocity $u^{\mu}$, the temperature $T$, the chemical potentials $\mu_{j}$, the shear stress tensor $\pi^{\mu \nu}$, and the bulk viscous pressure $\pi_{\mathrm{bulk}}$. We can now write the distribution function as an equilibrium part $f_{i,\text{eq}}$ plus a deviation from it, $\delta f_i$,
\begin{equation}
    f_i=f_{i,\text{eq}} + \delta f_i.
\end{equation}
The equilibrium distribution $f_{i,\text{eq}}$ can only depend on $T$, $u^\mu$ and $\mu_j$. The deviation $\delta f_i$ can in addition depend on $\pi^{\mu \nu}$ and $\pi_{\text{bulk}}$. For an ideal gas, where interactions are effectively unimportant, the ideal distribution function can be written as (we use $k_B=1$)
\begin{equation}
    f_{i,\text{eq}}(x,p)= \frac{1}{e^{\tfrac{-p_\mu u^\mu - \sum_j Q^i_j\mu_j}{T}j}\mp 1},
    \label{eq:feq}
\end{equation}
where the minus sign is for bosons and plus sign is for fermions, while $Q^i_j$ denotes the charge of the particle species $i$ with the associated chemical potential $\mu_j$. The non-equilibrium part, $\delta f_i$, is system-dependent \cite{Teaney:2003kp,Paquet:2015lta}, and not precisely known for the quark-gluon plasma. However, in our analysis we concentrate on the equilibrium part, $f_{i,\text{eq}}$, which gives the most important contribution to the phenomena we are after.

We consider now that, in full generality, it is possible to extend the Cooper-Frye prescription to two-particle (and similarly higher order) correlation functions. The distribution function in momentum space of pairs of particle type $j$ and $k$ is, similarly to Eq.~\eqref{eq:CF1P}, given by
\begin{equation}
E_p E_q \frac{dN_{jk}}{d^3 p d^3 q} = \frac{1}{(2\pi)^6} \int \limits_{\Sigma_{f}} d\Sigma_{\mu} d\Sigma^\prime_\nu \,  p^{\mu} q^\nu f_{jk}(x, x^\prime,p, q).
\label{eq:CFPairs}
\end{equation}
Factorization is often assumed (molecular chaos assumption), implying that the two-particle distribution function $f_{jk}(x, x^\prime,p, q)$ factorizes into the product of two single particle distributions, which would then also lead to a factorization of the left hand side of Eq.\ \eqref{eq:CFPairs}. This is, however, in conflict with the fluctuation-dissipation relation, see for example Ref.\ \cite{Calzetta:2008iqa}. In general,  we thus write
\begin{equation}
    f_{jk}(x,x^\prime,p,q)= f_j(x,p)f_k(x^\prime,q) + f_{jk}^{(\text{c})}(x,x^\prime,p,q),
    \label{eq:TwoParticleDistributionDecomposition}
\end{equation}
where $f_{jk}^{(\text{c})}$ denotes the connected part of the two-particle distribution in phase space. 

Let us comment on the meaning of fluctuations in the position of the freeze-out surface. What exactly determines $\Sigma_f$ is not precisely understood. The hypersurface is often characterized phenomenologically as the isothermal manifold where $T(x)=T_{\text{fo}}$, where $T_{\rm fo}$ is the \textit{freeze-out temperature} characterizing the transition from a QGP to a gas of hadrons, which occurs around $T_{\rm fo}=156$ MeV. Allowing for thermal fluctuations, in agreement with the fluctuation-dissipation relation, implies having a fluctuating $\Sigma_f$. However, for any practical implementation it is more convenient to work with a fixed $\Sigma_f$, and let the freeze-out temperature fluctuate. We take, thus, the freeze-out to be the manifold where the background temperature is fixed, $\bar T(x)= T_{\text{fo}}$, and we allow for a fluctuation of the actual temperature on this manifold, $T(x)=\bar T(x) + \delta T(x)$. For small fluctuations, specifically to linear order in $\delta T(x)$, this is equivalent to having a fluctuating surface $\Sigma_f$. We follow this prescription in the remainder of this manuscript.

\subsection{\label{sec:Symmetries of the freeze-out surface}Symmetries of the freeze-out surface and coordinates}

We are interested in particle production from high energy heavy-ion collisions in the central rapidity region \cite{Bjorken:1982qr}. There are two \textit{statistical} symmetries that can be used advantageously for central collisions (i.e., collisions at zero impact parameter), namely with respect to Bjorken boost and azimuthal rotations. These statistical symmetries manifest themselves in event averages and correlation functions. A coordinate choice that facilitates the use of these symmetries are the Milne coordinates. In the space of positions, they read
\begin{equation}
    \begin{aligned}
        x^{0} & = t = \tau \cosh(\eta)\, , & x^{1} = r\cos(\varphi) \\
        x^{2} & = r\sin(\varphi) \, ,        & x^{3} = z = \tau \sinh(\eta),
        \label{eq:Cartesian_param_coordinate}
    \end{aligned} 
\end{equation}
where we have introduced the Bjorken time, $\tau = \sqrt{t^2-z^2}$, the transverse radius $r = \sqrt{(x^{1})^2+(x^{2})^2}$, the azimuthal angle $\varphi = \arctan(x^{2}/x^{1})$, and spacetime rapidity $\eta=\mathrm{artanh}(z/t)$. Furthermore, we parametrize any curve in the $\tau-r$-plane via a parameter $\alpha$, that we may without loss of generality assume to be between 0 and 1. This implies in particular that the freeze-out surface can be written as
\begin{equation}
\begin{split}
\tau = \tau(\alpha),  \quad\quad\quad r = r(\alpha), \quad\quad\quad \alpha\in (0,1),\\
\varphi\in(0,2\pi), \quad\quad\quad  \eta\in (-\infty, \infty).
\end{split}
\end{equation}
Bjorken boosts and azimuthal rotations correspond to
\begin{equation}
    \begin{split}
        \varphi \rightarrow& \,\varphi^\prime = \varphi + \Delta \varphi, \\
        \eta \rightarrow& \,\eta^\prime = \eta + \Delta \eta,
        \label{eq:symmetryTransformations}
    \end{split}
\end{equation}
and we emphasize that event-averaged profiles in central collisions are invariant under such operations.
In the space of momenta, we introduce similar coordinates
\begin{equation}
    \begin{aligned}
        p^{0} & = E = m_{T}\cosh(y)=p^{\tau}, & p^{1} = p_{T}\cos(\phi), \\
        p^{2} & = p_{T}\sin(\phi),   & p^{3} = m_{T} \sinh(y),
    \end{aligned}
\end{equation}
with transverse momentum $p_{T}=\sqrt{(p^{1})^{2}+(p^{2})^{2}}$ and transverse mass $m_{T}=\sqrt{m^{2}+p_{T}^{2}}=\sqrt{E^{2}-(p^{3})^{2}}$. The momentum space rapidity $y$ and transverse angle $\phi$ are given by $y=\mathrm{arctanh}(p^{3}/p^0)$ and  $\phi=\textrm{arctan}(p^2/p^1)$, respectively.

\section{\label{sec:3} Background-fluctuation splitting and two-body correlations}

Relativistic fluid dynamics is a theory for the expectation values (defined with respect to some density operator) of the energy momentum tensor $\bar T^{\mu\nu}(x) \equiv \langle T^{\mu\nu} \rangle$ and other conserved quantities such as the net-baryon number current, $\bar N^\mu(x) \equiv \langle N^{\mu}(x) \rangle$. A given energy-momentum tensor and conserved current are decomposed as
\begin{equation}
\begin{split}
\bar T^{\mu\nu}= & \epsilon u^\mu u^\nu + (p+\pi_\text{bulk}) \Delta^{\mu\nu} + \pi^{\mu\nu}, \\
\bar N^\mu = & n u^\mu + \nu^\mu.
\end{split}
\label{eq:TmunuNmuExpectationValue}
\end{equation}
Here the fluid velocity is defined as the time-like eigenvector of the energy-momentum tensor, $\bar T^\mu_{\;\;\nu} u^\nu = - \epsilon u^\mu$, with eigenvalue given by energy density, $\epsilon$. In our conventions, the fluid velocity satisfies $u^\mu u_\mu = -1$. The projector orthogonal to the fluid velocity is given by $\Delta^{\mu\nu} = g^{\mu\nu} + u^\mu u^\nu$. The coefficient in front of it is decomposed into a thermal pressure, $p(\epsilon, n)$, related to $\epsilon$ and $n$ by the thermal equilibrium equation of state, and a the bulk viscous pressure, $\pi_\text{bulk}$. The shear stress, $\pi^{\mu\nu}$, is a symmetric and trace-less tensor that is orthogonal to the fluid velocity, $u_\mu  \pi^{\mu\nu}=0$. The conserved number current is parametrized by the fluid velocity $u^\mu$, the net particle number density, $n$, and the diffusion current, $\nu^\mu$, orthogonal to the fluid velocity, $\nu^\mu u_\mu=0$. 

The decomposition above associates a set of fluid fields to every  $\bar T^{\mu\nu}$ (with a time-like eigenvector) and $\bar N^\mu$, in a unique way (\textit{Landau matching}). Relativistic fluid dynamic does not, in a standard setting, make any statements about fluctuations around these expectation values. Extensions in this direction are however possible.

\subsection{Background-fluctution splitting}

In this work we consider a central heavy-ion collision, and perform a background-fluctuation splitting
\begin{equation}
\begin{split}
    T^{\mu\nu}(\alpha, \varphi,\eta)=& \bar T^{\mu\nu}(\alpha) + \Delta T^{\mu\nu}(\alpha, \varphi,\eta) ,\\ N^\mu(\alpha, \varphi,\eta) =& \bar N^\mu(\alpha) + \Delta N^\mu(\alpha, \varphi,\eta),
    \end{split}
    \label{eq:FluctuationTandN}
\end{equation}
where the background fields are the same as in Eq.\ \eqref{eq:TmunuNmuExpectationValue}. As anticipated, the background is invariant under Bjorken boosts and azimuthal rotations, and depends only on $\alpha$. We can construct, then, connected two-point correlation functions of the form
\begin{equation}
\begin{split}
\langle \Delta T^{\mu\nu}(x) \Delta T^{\rho\sigma}(x^\prime) \rangle,  \quad& \quad\quad \langle \Delta N^{\mu}(x) \Delta N^{\rho}(x^\prime) \rangle, \\
\langle \Delta T^{\mu\nu}(x) & \Delta N^{\rho}(x^\prime) \rangle,
\end{split}\label{eq:correlationsTmunuNmu}
\end{equation}
and note that the expectation values of the fluctuation fields vanish by construction. From the decompositions in Eq.~(\ref{eq:TmunuNmuExpectationValue}), we can express the correlation functions of Eq.~(\ref{eq:correlationsTmunuNmu}) terms of correlators $\langle \Delta \epsilon(x) \Delta \epsilon(x^\prime) \rangle$, $\langle \Delta u^\mu(x) \Delta u^\nu(x^\prime) \rangle $, etc., where $\Delta \epsilon(x)=\epsilon(x) - \Bar{\epsilon}(x)$, and so on. Here $\bar\epsilon(x)$ is the energy density that follows from the tensor decomposition (Landau matching) of the expectation value field in Eq.~\eqref{eq:TmunuNmuExpectationValue}. It is worth noting that one can replace the thermodynamic fluctuating variables using the equation of state.  For example, in the grand canonical ensemble, the energy density, $\epsilon$, and particle number density, $n$, are functions of temperature, $T$, and chemical potential, $\mu$. Such a change of variables can be done explicitly by considering the pressure $p(T,\mu)$, and invoking standard identities
\begin{equation}
    \epsilon=-p + T\frac{\partial p}{\partial T} + \mu \frac{\partial p}{\partial \mu}, \quad\quad n=\frac{\partial p}{\partial \mu}.
    \label{eq:StandardThermalIdentities}
\end{equation}
By linearizing these relations, we can reformulate all two-point functions in terms of $\langle \Delta T(x) \Delta T(y) \rangle$, $\langle \Delta T(x) \Delta \mu(y) \rangle$, and so forth. This defines the aforementioned fluctuating temperature and chemical potential in the present context. Working with fluid velocity, temperature and chemical potentials is advantageous as they determine the equilibrium distribution function in Eq.~\eqref{eq:feq}. 

Therefore, if we dub $\chi(x)$ any relevant fluctuation field in our description 
\begin{equation}
    \chi(x)=\left(\Delta T(x) , \Delta \mu(x) , \Delta u^\mu(x) , \Delta \pi^{\mu\nu}(x),\Delta \pi_{\text{Bulk}}(x) \right),
    \label{eq:NambuSpinor}
\end{equation}
we can translate fluctuations of conserved currents $T^{\mu\nu}$ and $N^\mu$ to two-point correlation functions $\langle \chi_s(x) \chi_t(x^\prime) \rangle$, where $s$ and $t$ denote two different fields. We relate now these correlators to the two-body particle distribution function at freeze-out introduced in Eq.~(\ref{eq:CFPairs}), and discuss the implications of such a relation on observables.

\subsection{\label{sec:Two point correlation function}Two-particle and two-point correlation functions}

Many observables can be constructed from the differential two-particle correlation function defined by
\begin{equation}
\label{eq: C_2 definition}
    C_{jk}(p,q)=\biggl \langle E_p E_q\frac{dN_{jk}}{d^{3}p d^{3}q} \biggr  \rangle - \biggl  \langle E_p\frac{dN_j}{d^{3}p} \biggr \rangle \biggl  \langle E_q\frac{dN_k}{d^{3}q} \biggr \rangle
\end{equation}
where $\langle \ldots \rangle$ denotes an average over events, and $j$, $k$ denote particle types. Equation ~\eqref{eq: C_2 definition} describes deviations of the pair distribution from a factorized form. In other words, it describes the relative contribution to the final particle spectra from effects beyond independent particle production. Note that, in the jargon of the hydrodynamic framework of heavy-ion collisions, independent particle production characterizes long-range \textit{flow} phenomena, in which particle correlations emerge solely from fluctuations of single-particle distributions \cite{Luzum:2011mm}. The feature we are after, i.e., the emergence of a genuine two-body contribution, $C_{jk}(p,q)\neq0$, represents, thus, a \textit{non-flow} phenomenon induced by thermal fluctuations at freeze-out.

The variation of the distribution function in Eq.~(\ref{eq:feq}) with respect to the fluctuation field can be characterized via a Taylor expansion
\begin{equation}
    \begin{split}
    \label{eq: distribution function product expansion}
   & f_j(x,p) = f_j(x,p) {\big |}_{0} + \sum \limits_{n}  \frac{\partial f_j(x,p)}{\partial \chi_{n}(x)}{\bigg |}_{0} \chi_n(x)\\
    &+ \frac{1}{2} \sum \limits_{mn}  \frac{\partial^2 f_j(x,p)}{\partial \chi_{n}(x) \partial \chi_m(x)} \bigg{|}_0 \chi_m(x)  \chi_n(x) + \ldots,
    \end{split}
\end{equation}
where the index $0$ denotes evaluation on the background level, i.e.\ for $\chi=0$. Considering now the product $f_j(x,p) f_k(x^\prime,q)$, taking the expectation value, and subtracting the product of expectation values, we obtain
\begin{align}
    \label{eq:averageddistributionfunctionproductexpansion}
    \nonumber  &\langle f_j(x,p)f_k(x^\prime,q)\rangle - \langle f_j(x,p)\rangle \langle f_k(x^\prime,q)\rangle  =\\ 
    &\sum \limits_{mn}\left\{ \frac{\partial f_j(x,p)}{\partial \chi_{m}(x)}\frac{\partial f_k(x^\prime,q)}{\partial \chi_n(x^\prime)}{\bigg |}_{0}\langle \chi_m(x) \chi_n(x^\prime) \rangle_\text{c} \right\} + \ldots~.
\end{align} 
In the notation of Eq.\ \eqref{eq:TwoParticleDistributionDecomposition}, the right hand side of Eq.\ \eqref{eq:averageddistributionfunctionproductexpansion} represents a contribution to the connected part of the two-particle phase-space distribution, $f^{(c)}_{jk}(x,x^\prime,p,q)$. To leading order in the fluctuation fields, then, the freeze-out integral in Eq.\ \eqref{eq:CFPairs} can be written as,
\begin{align}
\label{eq:relationtwoPointCorrelationFunctionTwoParticleCorrelation}
 \nonumber C_{jk}(p,q)=\frac{1}{(2\pi)^{6}} &\int_{\Sigma_f} d\Sigma_\mu d\Sigma^\prime_\nu \; p^\mu q^\nu  \\
 & \times \left[ \sum_{mn} \frac{\partial f_j}{\partial \chi_{m}}\frac{\partial f_k}{\partial \chi_n}{\bigg |}_{0}\langle \chi_m(x) \chi_n(x^\prime)\rangle_\text{c} \right].
\end{align}
This is the central result of this manuscript. It relates two-point correlation function of fluid fields on the freeze-out surface to two-body correlations in momentum space, of the form $E_pE_q\frac{dN_{\rm jk}}{d^3pd^3q}$, as introduced in Eq.~\eqref{eq: C_2 definition}. Before discussing the phenomenological consequences of this result, let us recast Eq.~(\ref{eq:relationtwoPointCorrelationFunctionTwoParticleCorrelation}) in a form which is more suitable to characterize observables.

\subsection{\label{sec:Two-particle mode expansion}Azimuthal and longitudinal mode expansion}

As a consequence of the statistical Bjorken and azimuthal symmetries of central collisions, the two-particle correlation function can only depend on the difference in rapidity and azimuthal angles. In other words
\begin{equation}
C_{jk}(p,q) = C_{jk}(p_T, q_T ,\Delta \phi, \Delta y),
\label{eq:twoPartCorrPTPhiEta}
\end{equation}
where $\Delta \phi=\phi_p - \phi_q$ and $\Delta y = y_p - y_q$. As often done in the analysis of particle spectra in heavy-ion collisions, we perform a Fourier decomposition of the pair distribution
\begin{equation}
\begin{split}
C_{jk}(p_T, q_T ; \Delta\phi, \Delta y) =& \int \frac{dk}{2\pi}\sum_{m=-\infty}^\infty e^{im\Delta\phi+ik \Delta y} \\
& \times c^{(m)}_{jk}(p_T, q_T ; k),
\end{split} \label{eq:twoPartCorrFluctuationExpansion}
\end{equation}
where $\Delta \phi=\phi - \phi^\prime$, $\Delta y = y-y^\prime$, $m$ and $k$ are, respectively, the azimuthal and rapidity wave numbers, and $c_{jk}^{(m)}$ is a differential Fourier coefficient in momentum space. The same can be done for the correlation functions of the freeze-out fluid fields, $\langle \chi_s(x) \chi_t(x^\prime)\rangle_c$. This quantity does depend solely on relative rapidity and azimuthal angle, $\Delta \varphi = \varphi - \varphi^\prime$ and $\Delta \eta = \eta - \eta^\prime$, such that it can also be decomposed in Fourier modes as
\begin{equation}
\begin{split}
\langle \chi_s(x) \chi_t(x^\prime)\rangle_c = & \int \frac{d k}{2\pi}\sum_{\tilde m=-\infty}^\infty  e^{i m \Delta\varphi+i k \Delta \eta} \\ & \quad\times g^{( m)}_{st}(\tau, \tau^\prime, r, r^\prime; k),
\label{eq:fluctuationCorrFourierExpansion}
\end{split}
\end{equation}
where $g^{( m)}_{st}(\tau, \tau^\prime, r, r^\prime; k)$ is now a differential Fourier coefficient in position space.

From \eqref{eq:relationtwoPointCorrelationFunctionTwoParticleCorrelation}, and using the freeze-out surface element 
\begin{equation}
\begin{split}
 -d\Sigma_\mu p^\mu = r(\alpha) \tau(\alpha) & { \Big [}\sqrt{m_j^2+p_T^2} \cosh(\eta-y)  \frac{dr}{d\alpha} \\
&  - p_T \cos(\varphi-\phi) \frac{d\tau}{d\alpha} {\Big ]} d\alpha d\varphi d \eta,
\label{eq:sigmap-contraction}
\end{split}
\end{equation}
we express the Fourier coefficients of the two-particle correlation function, $c^{(m)}_{jk}(p,q,k)$, as a function of the Fourier coefficient of the two-point functions of fluid fields, $g^{(m)}_{st}(\tau,\tau',r,r',k)$, on a mode-by-mode basis,
\begin{widetext}
\begin{equation}
\begin{split}
c^{(m)}_{jk}(p_T,q_T ; k) = & \frac{1}{(2\pi)^6}  \\
& \times \int d\alpha d\varphi d\eta \; r(\alpha) \tau(\alpha) { \Big [}\sqrt{m_j^2+p_T^2} \cosh(\eta)  r^\prime(\alpha)  - p_T \cos(\varphi) \tau^\prime(\alpha) {\Big ]} 
\\
& \times \int d\alpha^\prime d\varphi^\prime d\eta^\prime \; r(\alpha^\prime) \tau(\alpha^\prime) { \Big [}\sqrt{m_k^2+q_T^2} \cosh(\eta^\prime)  r^\prime(\alpha^\prime)  - q_T \cos(\varphi^\prime) \tau^\prime(\alpha^\prime) {\Big ]} 
\\
& \times \sum_{st} \frac{\partial f_j(p)}{\partial \chi_s} \frac{\partial f_k(q)}{\partial \chi_t} \; g^{(m)}_{st}(\tau, \tau^\prime, r, r^\prime; k) e^{i m\Delta \varphi+i k \Delta \eta}.
\end{split}\label{eq:cgRelation}
\end{equation}
\end{widetext}
This relation is a consequence of symmetries. Indeed, $m$ and $k$ label different representations with respect to azimuthal rotations and Bjorken boost transformations, and such representations cannot mix at the linear level.

Note that, to derive Eq.~\eqref{eq:cgRelation}, we have performed a shift in the integration variables $\eta-y \to \eta$, $\varphi-\phi\to \varphi$, and so forth. As a consequence, the combination $u_0^\nu p_\nu$ that enters the distribution function $f_j(p)$ becomes 
\begin{equation}
u_0^\nu p_\nu = - \sqrt{m_j^2+p_T^2} u_0^\tau(r) \cosh(y) + p_T u_0^r(r) \cos(\varphi),
\label{eq:up0-contraction}
\end{equation}
where we recall the only nonzero components of the background fluid velocity, $u_0^\mu$, are $u_0^\tau$ and $u_0^r$, which satisfy $(u_0^\tau(\alpha))^2-(u_0^r(\alpha))^2=1$.

\subsection{\label{sec:3c} Impact on correlation observables}

\subsubsection{Anisotropic non-flow coefficients}

Fourier harmonics are central to the phenomenology of heavy-ion collisions. We shall not focus on them in our phenomenological study, however, it is worth showing how they emerge in our formalism. Starting from the correlation function $C_{jk}(p_T,q_T,\Delta \phi, \Delta y)$, we expand in a Fourier series with respect to $\Delta \phi$. This leads to
\begin{align}
    \label{eq:vm defining expansion}
    & C_{jk}(p_{T},q_{T},\Delta \phi, \Delta y) = c^{(0)}_{jk}\lbrace 2 \rbrace ( p_{T},q_{T},\Delta y) \nonumber \\
    & + 2 \sum \limits_{m=1}^{\infty} c^{(m)}_{jk}\lbrace 2 \rbrace(p_T,q_T,\Delta y) \cos(m\Delta \phi).
\end{align}
The differential Fourier coefficients $c^{(m)}_{jk}\{2\}$ are then related to those appearing in Eq.~\eqref{eq:cgRelation} via
\begin{equation}
    c^{(m)}_{jk}\lbrace 2 \rbrace(p_{T}, q_{T},\Delta y)= \int \frac{dk}{2\pi} c^{(m)}_{jk}(p_T,q_T, k) e^{ik\Delta y},
    \label{eq:k-integrationOfCumulant}
\end{equation}
which leads to the following relation,
\begin{widetext}
\begin{equation}
    \begin{aligned}
    & c^{(m)}_{jk}\lbrace 2 \rbrace (k)= \frac{1}{(2\pi)^6}\int p_{T}dp_{T} \, q_{T}dq_{T}  \\
    & \times \int d\alpha d\varphi d\eta \; r(\alpha) \tau(\alpha) { \Big [}\sqrt{m_j^2+p_T^2} \cosh(\eta)  r^\prime(\alpha)  - p_T \cos(\varphi) \tau^\prime(\alpha) {\Big ]} 
    \\
    & \times \int d\alpha^\prime d\varphi^\prime d\eta^\prime \; r(\alpha^\prime) \tau(\alpha^\prime) { \Big [}\sqrt{m_k^2+q_T^2} \cosh(\eta^\prime)  r^\prime(\alpha^\prime)  - q_T \cos(\varphi^\prime) \tau^\prime(\alpha^\prime) {\Big ]} 
    \\
    & \times \sum_{s,t} \frac{\partial f_j(p)}{\partial \chi_s} \frac{\partial f_k(q)}{\partial \chi_t} \; g^{(m)}_{st}(\tau, \tau^\prime, r, r^\prime; k) e^{i m\Delta \varphi+i k \Delta \eta}.
    \label{eq:FullTwoPartCumulant}
    \end{aligned}
\end{equation}
\end{widetext}
With the inclusion of some appropriate normalization, this is nothing but a rapidity-differential version of the $p_T$-integrated variance of the $m$th order flow coefficient, which can be measured experimentally via 2-particle angular correlations. This result represents however the fluctuation of a \textit{non-flow} coefficient, arising from thermal fluctuations on the freeze-out surface. One drawback of such an observable is that the influence of thermal fluctuations on it is hard to detect in practice. Correlations induced by thermal fluctuations are inherently of short-range nature. As such, in an experiment they are buried under a large background of short-range two-body momentum correlations from, e.g., jets and resonance decays. As we shall argue in the following, this issue may however be solved if one looks at the angular correlations of fluctuations of conserved quantities, which we can also quantify in our framework.

\subsubsection{\label{sec:332} Net-charge correlations}

Fluctuations of conserved quantities have been discussed in the context of high-energy nuclear physics since the early days of the heavy-ion collision program. In the grand canonical ensemble, at fixed chemical potential, one can study local fluctuations in net baryon number. The associated moments and cumulants have been proposed as potential probes of the QCD phase diagram, notably, of potential critical phenomena at the QGP-hadron transition  \cite{Stephanov:1998dy}. Such critical phenomena may leave an imprint in the particle distribution not only via their fluctuations at a given QGP volume and rapidity acceptance, but also within different momentum and azimuthal angle bins. Although the impact of statistical fluctuations and conservation laws in the process of particlization of the QGP have been discussed at length in the literature (see e.g. \cite{Schwarz:2017bdg,Oliinychenko:2019zfk,Oliinychenko:2020cmr,Vovchenko:2021yen} for recent studies), to the best of our knowledge net-charge fluctuations via a differential observable such as the two-particle correlation function in Eq.~(\ref{eq: C_2 definition}) have not been studied yet. 

The formalism worked out in the previous section can be straightforwardly applied in such an endeavour. We consider, as an example, the event-by-event net-baryon number, $N_B=N_b-N_{\bar b}$, where $N_b$ is the number of baryons, while $N_{\bar b}$ is the number of anti-baryons. At freeze-out (or QGP-hadron transition), an analysis of fluctuations of the net-baryon number can be achieved by looking at the the following correlation function
\begin{equation}
\begin{split}
    &\left \langle E_p\frac{dN_B}{d^3p}E_q\frac{dN_B}{d^3q} \right\rangle =\\
    &\left\langle \left( E_p\frac{dN_\mathrm{b}}{d^3p}-E_p\frac{dN_{\mathrm{\bar{b}}}}{d^3p}\right)\left( E_q\frac{dN_{\mathrm{b}}}{d^3q}-E_q\frac{dN_{\mathrm{\bar{b}}}}{d^3q}\right) \right\rangle.
    \end{split}
\end{equation}
Dubbing $B_j$ the baryon number of hadronic species $j$, we can rewrite that expression as a weighted sum over two-particle correlation functions,
\begin{equation}
    \left \langle E_p\frac{dN_B}{d^3p}E_q\frac{dN_B}{d^3q} \right\rangle = \sum \limits_{jk} B_j B_k \left \langle E_p E_q\frac{dN_{jk}}{d^3p d^3q}\right \rangle_c.
    \label{eq:Netbaryon NumberMomentumSpace}
\end{equation}
This can be generalized to any other conserved charges, such as electric charge, strangeness, or heavy quark numbers. The quantity in the bracket of the right-hand side of Eq.~(\ref{eq:Netbaryon NumberMomentumSpace}) can, hence, be related to the correlation functions of the thermodynamic fields on the freeze-out surface, as discussed in the previous sections. One interesting outcome of such procedure, which we shall emphasize in Sec.~\ref{sec:5}, is the possibility of quantifying net-charge correlations coming from variances of the chemical potentials, $\langle \mu_{B,Q,S}~ \mu_{B,Q,S} \rangle$, that emerge when the expectation values of the chemical potentials vanish. This opens a window to look for dynamical net-charge fluctuations at very high-energy colliders. In the next section, Sec.~\ref{sec:4}, we build a model of the fluctuating kinetic freeze-out surface that will allow us, in Sec.~\ref{sec:5}, to quantify such phenomena with a numerical calculation.

\section{Model of fluctuating kinetic freeze-out surface}
\label{sec:4}

We construct now a model of the fluctuating kinetic freeze-out surface of central heavy-ion collisions, which we will employ later on in Sec.~\ref{sec:5} to study numerically the observable consequences of our findings. Our model requires three ingredients, namely, the average particle spectra (1-point functions), the fluctuating fields that determine two-particle correlations in the final state (2-point functions), and an equation of state to know how different fields are related among each other.

\subsection{\label{sec:Explicit freeze-out surface and the blast wave model}One-point functions: blast wave parametrization}

To keep our calculation as analytically-controlled as possible, we resort to a very simple description of the kinetic freeze-out surface. We employ the blast-wave parametrization \cite{Schnedermann:1993ws}, corresponding to a freeze-out at constant proper time $\tau$. The freeze-out surface is in this case parametrized by
\begin{equation}
\label{eq:kfo}
\begin{split}
\tau = \tau_\text{fo}, & \quad\quad\quad r \in [0,  R_\text{fo}], \\
\varphi\in [0,2\pi), & \quad\quad\quad \eta\in (-\infty, \infty),
\end{split}
\end{equation}
along with a fluid velocity which we parametrize via the blast-wave profile 
\begin{equation}    
\label{eq:BWvelo}
v^r_0(r) = \frac{u^r_0(r)}{u^\tau_0(r)} = \beta_{S} \left( \frac{r}{R_{\mathrm{fo}}}\right)^{n}.
\end{equation}
In this model, then, the  momentum spectrum averaged in a class of events is, for hadron species $j$ and in the Boltzmann approximation, given by (with $\alpha=r/R_{\rm fo}$)
\begin{align}
          \nonumber & \left\langle E_p\frac{dN_j}{d^{3}p}\right\rangle = %\frac{dV}{dy}
          \tau_\text{fo} R_\text{fo}^2
          \frac{\nu_j}{2\pi^{2}} \exp\left( \sum_k Q_k(j) \frac{\mu_{k,0}}{T_\text{fo}}\right) \sqrt{m_k^2+p_T^2} \\
        & \times \int \limits_{0}^{1} d\alpha \, \alpha \; K_{1}\left[\frac{\sqrt{m_j^2+p_T^2} u_{0}^{\tau}(\alpha)}{T_\text{fo}}\right] I_{0}\left[\frac{p_{T}u_{0}^{r}(\alpha)}{T_\text{fo}}\right].
        \label{eq:Blast_Wave_Boltzmann}
\end{align}
In this equation, $\nu_j$ is the spin degeneracy, $Q_k(j)$ is the value of charge $k$ for species $j$, where $k=B$, $Q$, or $S$, for baryon number, electric charge or strangeness,  and $\mu_k$ is the background chemical potential associated with charge $k$. Also, $T_{\rm fo}$ is the background temperature, $m_j$ is the hadron mass, and $K_1$ and $I_0$ are the modified Bessel functions. Note that the formula does not involve $\tau_{\rm fo}$ and $R_{\rm fo}$ introduced in Eq.~(\ref{eq:kfo}), but only the product $\tau_{\rm fo} \pi R_{\rm ro}^2$, which we identify with a freeze-out volume per unit rapidity, $dV/dy$.

\subsection{Two-point functions: thermal correlations in an ideal gas}

\label{sec:4b}

For the two-point correlation functions of the thermodynamic fields at freeze-out, we consider the simple case of an ideal gas. This implies, notably, that the two-point correlators are local , i.e.,
\begin{equation}
\label{eq:pref}
    \langle \chi_n(\alpha) \chi_m(\alpha')\rangle \propto \delta^{(3)} (\alpha-\alpha'),
\end{equation}
where we recall that $n$ and $m$ label two different thermodynamic fields, and that the correlation functions depend only on $\alpha$. Working within the framework of classical statistical field theory, we are able to evaluate the prefactors of Eq.~(\ref{eq:pref}), for any choice of the thermodynamic fields. We describe now the salient points of the derivation, and we refer to Appendix~\ref{App:A} for a detailed description of our calculations.

The starting point is a probability of fluctuations for fields $\chi$ defined on the freeze-out surface, $\Sigma$. This has some generic form 
\begin{equation}
\label{eq:1}
p[\chi] = \frac{1}{Z_\Sigma} e^{-I_\Sigma[\chi]},
\end{equation}
where the partition function, $Z_\Sigma$, carries all the information about the correlations of $\chi$ fields, namely,
\begin{equation}
\begin{split}
& \langle \chi_n(\alpha) \chi_m(\beta) \rangle = \\
& \frac{1}{Z_\Sigma [J]} \left( \frac{1}{\sqrt{h(\alpha)}} \frac{\delta}{\delta J_n(\alpha)} \right) \left( \frac{1}{\sqrt{h(\beta)}} \frac{\delta}{\delta J_m(\beta)} \right) Z_\Sigma [J]{\big |}_{J=0},
\end{split}
\label{eq:3}
\end{equation}
where $J_n$ is an external source associated with the field $\chi_n$.  Now, under a strong local equilibrium hypothesis, where all fluid cells are in thermal equilibrium and all fluctuating independently, one can show that the functional $I_{\Sigma[\chi]}$ is quadratic in the fields $\Delta T$, $\Delta \mu_j$, and the three independent components of $\Delta u^\mu$, thus yielding, somewhat unsurprisingly, a Gaussian probability distribution for their fluctuations. As a consequence, one can evaluate explicitly the partition function, $Z_{\Sigma}$, and the correlators of $\chi$ fields. The local nature of the probability functional is a consequence of the strong local equilibrium assumption, and it leads to correlations of the form
\begin{equation}
\langle \chi_n(\alpha) \chi_m(\alpha^\prime) \rangle_c = \frac{1}{\sqrt{h(\alpha)}} \delta^{(3)}(\alpha-\alpha^\prime) \sigma_{nm}(\alpha).
\label{eq:finalChi}
\end{equation}
The matrix of variances, $\sigma_{nm}$, and its inverse can be derived analytically. For the choice $\chi_n=(\Delta T, \Delta\mu_i, \Delta u^\rho)$,  it reads ($n^\mu$ is the vector normal to the surface element) 
\begin{equation}
\sigma^{-1} = \begin{pmatrix} - \frac{n\cdot u}{T} \frac{\partial^2 p}{\partial T^2} & -\frac{n\cdot u}{T} \frac{\partial^2 p}{\partial T \partial \mu_j} & - \frac{n_\sigma}{T}\frac{\partial p}{\partial T} \\ 
- \frac{n\cdot u}{T} \frac{\partial^2 p}{\partial T \partial\mu_i} & - \frac{n \cdot u}{T} \frac{\partial^2 p}{\partial\mu_i \partial \mu_j} & - \frac{n_\sigma}{T} \frac{\partial p}{\partial \mu_j} \\ - \frac{n_\rho}{T} \frac{\partial p}{\partial T} & - \frac{n_\rho}{T} \frac{\partial p}{\partial \mu_i} & - \frac{n\cdot u}{T}(\epsilon+p) \Delta_{\rho\sigma}
\label{eq:Matrix}
\end{pmatrix}.
\end{equation}
At this stage, then, we are able to evaluate the correlators of $\chi$ fields, which can be plugged into Eq.~(\ref{eq:relationtwoPointCorrelationFunctionTwoParticleCorrelation}), to yield two-body correlation functions in momentum space, $\bigl \langle E_p E_q \frac{dN_{ij}}{d^3pd^3q}  \bigr\rangle$. The only missing ingredient is the equation of state, which specifies the relation between the thermodynamic fields, and which we discuss now.

\subsection{\label{sec:Thermodynamic equation of state}Thermodynamic equation of state}

Many features of the thermodynamic equation of state describing a system at the intersection of a QGP and a gas of hadrons, as it occurs around freeze-out, are captured well by the equation of state of the free hadron resonance gas (HRG) \cite{Vovchenko:2014pka},
\begin{equation}
    \begin{aligned}
        p_{\mathrm{HRG}} & = \sum_{j}\frac{d_{j}}{2\pi^2} \int dm \; b_{j}(m)  \sum \limits_{\kappa = 1}^{\infty} T^2 m^2 \frac{s_{B/F}^{\kappa+1}}{\kappa^2} \\ 
        & \times \exp\biggl (\frac{\sum \limits_{i} Q_i(j)\mu_i}{T}\kappa \biggr) K_{2}\left(\frac{m}{T}\kappa\right),
    \end{aligned}
    \label{eq:eos_hrg_protons}
\end{equation}
where $s_{B/F} =1$ for bosons, $s_{B/F} =-1$ fermions and $b_{j}(m)$ describes the width of the resonance $j$. For stable particles $b_{j}(m)=\delta(m-m_j)$. Note that, in a practical implementation, the sum involving the variable $\kappa$ can be truncated after the first $\approx 5$ terms to get an excellent convergence for all hadron species. 

This completes our realization of a fluctuating kinetic freeze-out surface in high-energy heavy-ion collisions. We move on, then, to a numerical application of this model.

\section{\label{sec:5} Numerical study}

Let us make a brief recap of the approximations used in the construction of our model. As it stands, Eq.\ \eqref{eq:relationtwoPointCorrelationFunctionTwoParticleCorrelation} determines the differential two-particle correlation function from correlation functions of the fluid fields in Eq.\ \eqref{eq:NambuSpinor}, which are given by Eq.~(\ref{eq:finalChi}). Our derivation is based on arguments of a thermodynamic kind, which can be used in the context of interacting theories and which follow mainly from conservation laws and entropy maximisation. We assume conservation of energy and momentum, as well as quantum numbers like baryon or electric charge. Around kinetic freeze-out, when free streaming sets in, there are however additional conservation laws due to the absence of scatterings. Therefore, we expect our formalism to give a reasonable description of correlation functions of conserved quantities like energy, momentum, and conserved charges. Correlation functions of individual identified particles, which are outside the reach of standard QCD thermodynamics, should instead be taken with a grain of salt. For this reason, in this section we focus on the correlation functions of net charges (baryon and electric charge), as discussed in Sec.~\ref{sec:3}.

Concerning our implementation, for the blast-wave parametrization of the freeze-out surface we use the parameters returned by the fit of Ref.~\cite{Mazeliauskas:2019ifr} for Pb+Pb collisions at $\sqrt{s_{\rm NN}}=2.76$ TeV in the 0-5\% centrality class, corresponding to $n=0.34$ and $\langle \beta_T \rangle = \frac{2\beta_s}{2+n}=0.66$ in Eq.~(\ref{eq:BWvelo}), as well as $\frac{dV}{dy}=4273$ fm$^3$ and $T_{\rm fo}=149$ MeV in Eq.~(\ref{eq:Blast_Wave_Boltzmann}). For collisions at such high energies, we set $\mu_B=\mu_Q=\mu_S=0$ at the background level, though we emphasize the variances of these quantities will be nonzero due to thermal fluctuations. We take into account charged pions and kaons, as well as protons and anitprotons. The masses, charges and degeneracies of these particles are taken from the PDG \cite{ParticleDataGroup:2020ssz}. These are also used in the HRG equation of state.

Additionally, we evaluate the two-particle correlation function $C_{jk}(p_T,q_T,\Delta \phi, \Delta y)$ from its Fourier-expanded expression given in Eq.~(\ref{eq:twoPartCorrFluctuationExpansion}). This requires calculating the Fourier coefficients $g_{st}^{(m)}$ appearing in Eq.~(\ref{eq:cgRelation}).
With the freeze-out surface now clearly defined, these modes are given by the Fourier transform of Eq.\ \eqref{eq:finalChi} with respect to $\varphi$ and $y$, which simply yields
\begin{equation}
    g_{st}^{(m)}(\tau,\tau^{\prime}, r,r^{\prime};k)=
    \frac{1}{2\pi} \frac{1}{\sqrt{h(\alpha)}} \delta(\alpha-\alpha^{\prime})\sigma_{st}(\alpha),
    \label{eq:gFourierMode}
\end{equation}
where  $\sigma_{st}(\alpha)$ is the same thermal correlation matrix of Eq.~(\ref{eq:Matrix}). Note that the r.h.s. of Eq.~\eqref{eq:gFourierMode} is independent of $m$ and $k$, as a consequence of the correlation function with vanishing range in Eq.\ \eqref{eq:finalChi}. In the following we keep working with an ideal fluid, and consider the following fluctuation fields, $\Delta T$, $\Delta \mu_k$ for $k=B$, $Q$, $S$, $\Delta u^\mu$ for three independent components, such that the matrix of variances $\sigma_{st}$ has $7 \times 7$ entries.

\subsection{\label{sec:Net baryon fluctuations.}Net-charge correlations}
\begin{figure*}[t]
    \centering
    \includegraphics[width=\linewidth]{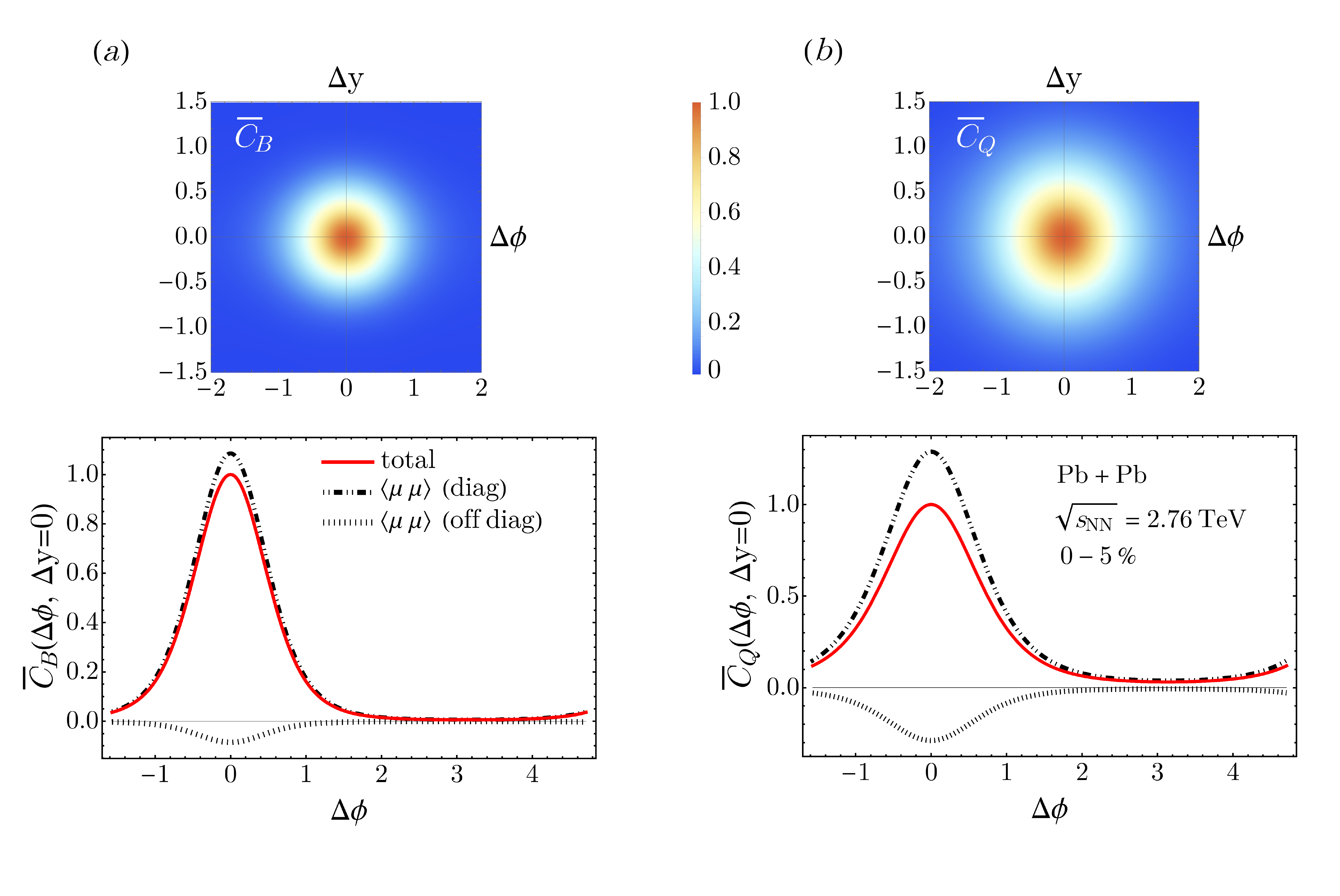}
    \caption{Differential two-particle net-charge correlation, normalized to unity at their peak, as a function of relative azimuthal angle, $\Delta \phi$, and relative momentum rapidity, $\Delta y$. (a): net-baryon correlations, $\bar C_B$. (b): net-charge correlations, $\bar C_Q$. The upper panels show two-dimensional profiles, while the lower panels are at fixed $\Delta y=0$. We note that all the contributions from field correlations that are not shown in the graphs, $\langle \chi_s \chi_t \rangle$, where both $s$ and $t$ do not represent chemical potentials, are zero.}
    \label{fig:1}
\end{figure*}

We recall the weighted sum in Eq.~(\ref{eq:Netbaryon NumberMomentumSpace}) defining the two-particle correlation for the net-baryon number:
\begin{equation}
    \left \langle E_p\frac{dN_B}{d^3p}E_q\frac{dN_B}{d^3q} \right\rangle = \sum \limits_{jk} B_j B_k \left \langle E_p E_q\frac{dN_{jk}}{d^3p d^3q}\right \rangle.
%    \label{eq:Netbaryon NumberMomentumSpace}
\end{equation}
We compute this quantity from Eq.~(\ref{eq:relationtwoPointCorrelationFunctionTwoParticleCorrelation}) in our framework by considering the contribution of protons and anti-protons, i.e., by taking the common assumption that net-baryon fluctuations correspond to net-proton fluctuations. We integrate Eq.~(\ref{eq:cgRelation}) up to $m=10$ and $k=10$, which guarantees convergence, to exhibit the angular dependence of the two-particle correlation, 
\begin{equation}
C_B (\Delta \phi, \Delta y) \equiv \left \langle E_p\frac{dN_B}{d^3p}E_q\frac{dN_B}{d^3q} \right\rangle (\Delta \phi, \Delta y).
\end{equation}
For simplicity we normalize our correlation function in such a way that it is unity at $\Delta \phi = \Delta y = 0$,
\begin{equation}
 \bar     C_B (\Delta \phi, \Delta y) = C_B (\Delta \phi, \Delta y) / C_B (0, 0).
\end{equation}
This allows us to point out generic features of the \textit{shape} of the correlation function. The normalization can be chosen, then, depending on what is most convenient to measure in experiments. Normalizing the correlation function with the average net baryon charge would lead to a division by very small numbers, especially at ultrarelativistic collision energies. To avoid such issues, one possibility is, for instance, to normalize with standard deviations (which are non-vanishing), and thus construct Pearson correlation coefficients.

Our result for $\bar C_B$ is displayed in Fig.~\ref{fig:1}(a). It gives a good illustration of the kind of influence thermal fluctuations have on two-particle correlations. 

The correlation is indeed short-range, and leads to a broadening of the near-side peak. The diffusion of the near-side peaks is in fact quite sizable, as the peak drops to half its maximum around the circle $\Delta y = \Delta \phi = 1 $. We take now a slice of the two-dimension profile corresponding to $\bar C_B(\Delta \phi, \Delta y = 0)$, shown as a red solid line in the bottom panel of Fig.~\ref{fig:1}(a). Much insight can be gained by decomposing this quantity into contributions coming from different correlators of fluid fields. Since we are looking at a fluctuation of conserved charges that can only be engendered by a fluctuation in the corresponding chemical potentials, we find that the contribution to $\bar C_B(\Delta \phi, \Delta y = 0)$ coming from the variances of the form $\langle \chi_s\chi_t \rangle$, where both $s$ and $t$ do not represent a chemical potential, is zero. The correlation of the net-baryon number is, thus, produced solely by diagonal and off-diagonal variances of the form $\langle \mu_{B,Q,S} \mu_{B,Q,S} \rangle$. Naturally, in our case of net-proton correlations, the contribution of $\langle \mu_S \mu_S \rangle$ does vanish, as a fluctuation in $\mu_S$ does not change the proton spectrum. The total peak in the bottom panel of Fig.~\ref{fig:1}(a) gets, thus, a dominant contribution from the diagonal variances $\langle \mu_B \mu_B \rangle$ and $\langle \mu_Q \mu_Q \rangle$, and a negative subleading, albeit visible contribution from the off-diagonal one, $\langle \mu_Q \mu_B \rangle$.

In Fig.~\ref{fig:1}(b), we perform the same analysis albeit focusing on the net electric charge, $ \bar C_Q(\Delta \phi, \Delta y)$, which we construct from pions, kaons, protons and their antiparticles. The same features are found, though the resulting near-side peak is visibly more diffuse compared to $\bar C_B$.

We comment now on the implications of this result. One could simply take the particles collected in central Pb+Pb collisions at top Large Hadron Collider energy, and measure the correlation functions depicted in Fig.~\ref{fig:1}. The question is, hence, to which extent one should expect the experimental result to resemble our theoretical results. While care is needed, there is no obvious reason to believe the experimental results will look entirely different from ours. %The point is precisely that, besides thermal fluctuations, there are essentially no other phenomena leading to the fluctuation of conserved charges. The plethora of non-flow correlations affecting the near-side peak, e.g. jet and resonance decays, do strictly conserve the baryon number and the electric charge. 
Within a rapidity window that is large enough to detect all of those decay products, one does not in general expect non-flow phenomena to lead to fluctuations of conserved charges. This may also be investigated by comparing the experimental results to the output of event-generators that include non-flow contributions. Resonance decays may also be added in future to our formalism by modifying the blast-wave parametrization of Eq.~(\ref{eq:Blast_Wave_Boltzmann}), as precisely done in Ref.~\cite{Mazeliauskas:2019ifr}.  That said, a potential modification of the correlations measured experimentally could however come from self-correlations. Experiments construct their correlation observables from pairs of distinct particles, while this is not something that we require in our calculations. We discuss this subtle point in more depth in Sec.~\ref{sec:5c}.

\subsection{\label{subsec:Introducing a finite correlation length}Introducing a finite correlation length}

We see that already the use of ultra-local thermal correlations in Eq~(\ref{eq:finalChi}) leads to a significant spread of the near-side peaks in Fig.~\ref{fig:1}. This suggests that this spread may be effectively independent of whether correlations are local or not. In this subsection, we perform a simple calculation to check this explicitly. We go beyond the ultra-local approximation by replacing the delta functions with exponential decays, characterized, in both $\alpha$, $\varphi$, and $\eta$, by correlation lengths $\xi_\alpha$, $\xi_\varphi$, $\xi_\eta$, respectively, while keeping the same correlation matrix $\sigma_{st}$ as prefactor. Working for convenience with the variables $(\alpha+\alpha')/2$ and $\alpha-\alpha'$, we have in short:
\begin{align}
 \nonumber \langle \chi_s(x) \chi_t(x^\prime) \rangle &= \frac{1}{\sqrt{h(\frac{\alpha+\alpha'}{2})}}  \sigma_{st}\biggl(\frac{\alpha+\alpha'}{2}\biggr) \\
  &\times\frac{1}{2\xi_\alpha} e^{-\tfrac{|\alpha-\alpha^\prime|}{\xi_\alpha}} \frac{1}{2\xi_\varphi} e^{-\tfrac{|\varphi-\varphi^\prime|}{\xi_\varphi}}  \frac{1}{2\xi_\eta} e^{-\tfrac{|\eta-\eta^\prime|}{\xi_\eta}}.
  \label{eq:FiniteCorrExpDecay}
\end{align}
In the calculation of, e.g., $C_{B}(\Delta \phi, \Delta y)$, the part that gets modified is the Fourier coefficient $g_{st}$ in Eq.~(\ref{eq:gFourierMode}). This requires, once again, to perform Fourier transforms with respect to $\Delta \varphi=\varphi-\varphi'$ and $\Delta \eta = \eta - \eta'$, i.e.,
\begin{equation}
    \int_{-\pi}^{\pi} \frac{d\Delta \varphi}{2\pi} \,\frac{1}{2\xi_\varphi} e^{-\tfrac{|\Delta \varphi|}{\xi_\varphi}-im\Delta \varphi}\int_{-\infty}^{\infty}d\Delta \eta\frac{1}{2\xi_\eta}e^{-\tfrac{|\Delta \eta|}{\xi_\eta}-ik\Delta \eta},
\end{equation}
which simply yields
\begin{equation}
\begin{split}
    g_{st}^{(m)}(\alpha,\alpha^\prime;k)=& \frac{1}{2\pi} \frac{1}{\sqrt{h(\frac{\alpha+\alpha'}{2})}} \sigma_{st}\left(\tfrac{\alpha + \alpha^\prime}{2}\right)  \frac{1}{2\xi_\alpha}e^{\tfrac{-| \alpha-\alpha'|}{\xi_\alpha}}\\
    \times& \frac{1}{1+\xi_\eta^2 k^2} \frac{1 -  e^{-\tfrac{\pi}{\xi_\varphi}}\cos(m\pi)}{1 + \xi_\varphi^2 m^2},
    \label{eq:gFiniteCorrLength}
    \end{split}
\end{equation}
that depends now on both $m$ and $k$.

\begin{figure}[t]
    \centering
    \includegraphics[width=\linewidth]{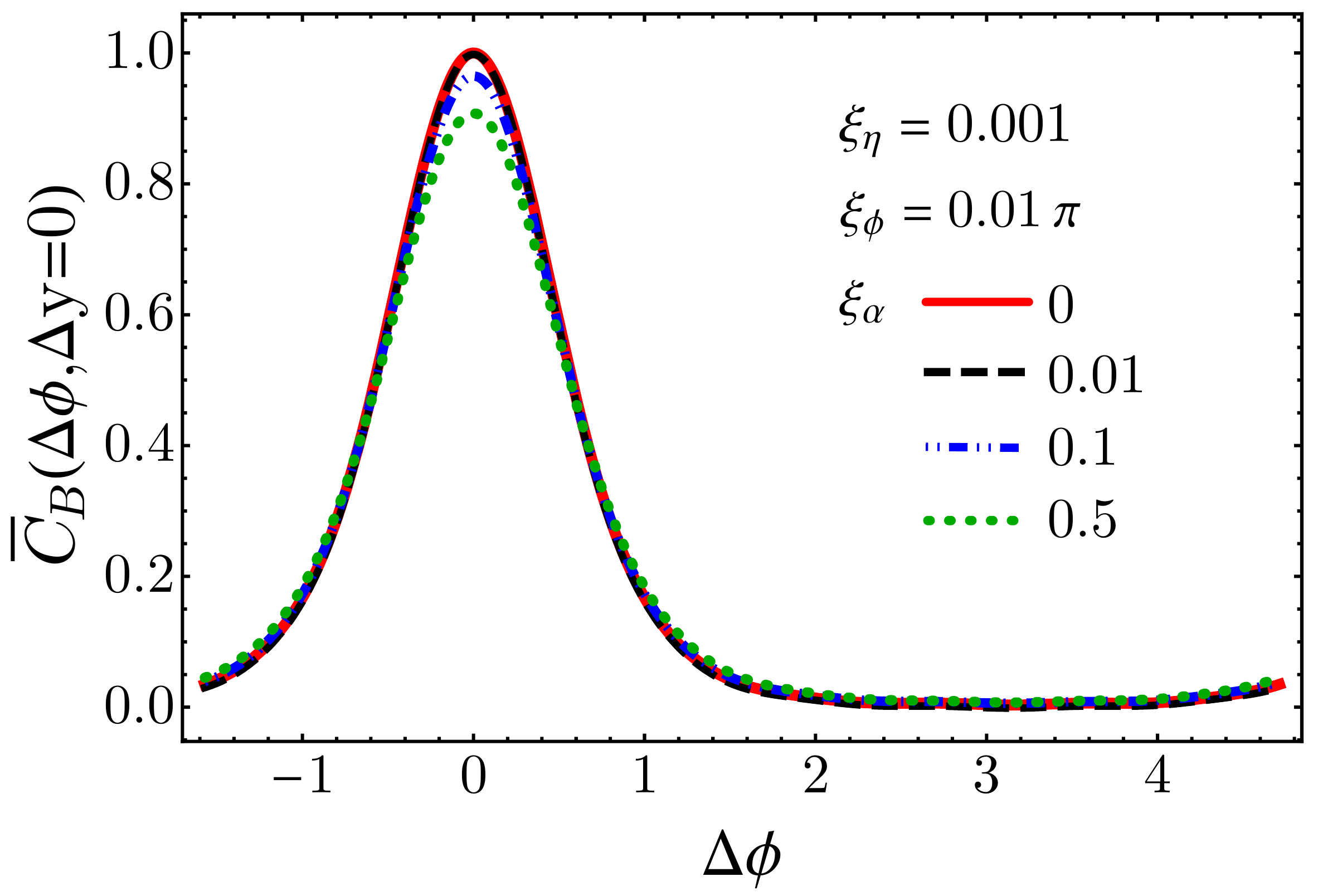}%
    \caption{Same as in the lower panel of Fig.~\ref{fig:1}(a), but with varying correlation length (shown as different line styles) along the $\alpha$ direction, following Eq.~(\ref{eq:FiniteCorrExpDecay}).}
    \label{fig:2}
\end{figure}

For a numerical application, we fix $\xi_\varphi=0.01\pi$ and $\xi_\eta=0.001$, and we vary $\xi_\alpha$. The resulting correlation function, $\bar C_B(\Delta \phi, \Delta y=0)$, is plotted in Fig.~\ref{fig:2}. Unsurprisingly, increasing the correlation length along $\alpha$ over a wide interval does yield only a minor additional diffusion of the near-side peak. While this prescription is rather \textit{ad hoc}, it suggests that the precise shape of the thermal correlators in Eq.~(\ref{eq:finalChi}) may indeed not matter much for the final result. The more crucial ingredient is the thermal correlation matrix.

\subsection{\label{sec:5c}Self-correlations}

Before concluding, let us comment on the issue of self-correlations. Two-particle correlation functions as Eq.~\eqref{eq: C_2 definition} can be estimated experimentally in two different ways. The choice is whether the two correlated particles are distinct, or whether the same particle can form a pair with itself. Typically, as done in standard flow analyses, only pairs of distinct particles are chosen, to avoid a strong bias in the resulting correlations. As it stands, the formalism developed in Secs.~\ref{sec:3} and \ref{sec:4b} features self-correlations. This can be seen in the limit where the thermodynamic equation of state entering the functional $I_{\Sigma[\chi]}$ in \eqref{eq:1} is the one for an ideal gas, and non-trivial correlations are in fact of the self-correlation type.

To see this more explicitly, for a non-interacting gas of free streaming particles, one can write at $x^0=x^{\prime 0}$ \cite{Calzetta:2008iqa}
\begin{equation}
\begin{split}
& \langle \Delta f_j(x,p) \Delta f_k(x^\prime, q) \rangle = \\ & \delta_{jk} (2\pi)^3 \delta^{(3)}(\mathbf{x}-\mathbf{x}^\prime) \delta^{(3)}(\mathbf{p}-\mathbf{q}) f_{\text{eq}}(x,p)(1\pm f_{\text{eq}}(x,p)),
\end{split}
    \label{eq:FluctuationsFeq}
\end{equation}
where the upper (lower) sign is for bosons (fermions), or replace $(1\pm f_\text{eq}) \to 1$ for Boltzmann statistics. One should understand the right-hand side of Eq.\ \eqref{eq:FluctuationsFeq} as a contribution to the connected pair correlation function in phase space $f_{jk}^{(\text{c})}(x,x^\prime,p,q)$ in Eq.\ \eqref{eq:TwoParticleDistributionDecomposition}, and, accordingly, it contributes in the freeze-out integral in Eq.\ \eqref{eq:CFPairs}. This is a contribution from self-correlations. On the other hand, one may take integrals over positions and momenta in Eq.~\eqref{eq:FluctuationsFeq} and obtain the second order cumulants of thermodynamic variables, such as energy or conserved particle numbers, as they follow for an ideal (quantum) gas. These arguments show in which sense our formalism reduces to self correlations in the limit of an ideal gas. 

Let us stress that existing literature describing hydrodynamic fluctuations in heavy-ion collisions has also been dealing with this question, see e.g. the recent Refs.~\cite{Kapusta:2017hfi,De:2020yyx}. Fluctuations induced by hydrodynamic noise are evolved in time dynamically, and self-correlations are isolated. Their impact is quantified on measurements on balance functions \cite{Pratt:2012dz}, and it is indeed a significant correction. In our framework, it might also be possible to subtract self-correlations by modifying the correlation functions on the freeze-out surface. For instance, one could devise an analogue of factorial cumulants \cite{Kitazawa:2017ljq}, that remove self-correlations by construction. Devising such a subtraction scheme goes beyond the scope of this work, such that our framework has self-correlations included. Future experimental measurements of the quantities proposed in Fig.~\ref{fig:1} should probably be performed both with and without self-correlations, to assess their effect.

\section{\label{sec:6} Conclusion and outlook}

In this manuscript we have addressed a rather straightforward problem. Given a surface $\Sigma$ characterized by a set of fluid fields, we have studied position-space correlations among such fields induced by thermal fluctuations on $\Sigma$. Identifying $\Sigma$ with the kinetic freeze-out surface of heavy-ion collisions, we have then translated such correlations into momentum-space correlations at the particlization of such surface following the Cooper-Frye formula generalized to account for genuine two-body contributions. A straightforward application of our formulas combining a blast-wave parametrization of the average momentum yields with the thermal correlation functions of an ideal resonance gas leads to the emergence of angular net-charge correlations from thermal fluctuations at the freeze-out surface. An interesting diffusion of the near-side peak is observed, which we expect may motivate further experimental and theoretical developments.

Experimentally, the proposed \textit{non-flow} correlations can be measured straightforwardly. The long-range contribution from the hydrodynamic flow can be removed from the near-side peak by fitting its shape at the away-side peak. For the correlatons of net charges, it would be interesting, in particular, to compare experimental data with the output of traditional event generators, which do not include potential effects of \textit{dynamical} fluctuations. This can be done both at top LHC energy where $\mu_{B,Q,S}=0$, as well as a function of beam energy at RHIC, where our results could be generalized to include finite chemical potentials (albeit with a background freeze-out surface that is no longer boost-invariant). 

On the theory side,  our calculations can certainly be improved in several ways, by, e.g., adding viscous corrections at freeze-out, relaxing the strong equilibrium hypothesis, as well as implementing
an equation of state allowing for a more sensible study of critical phenomena. The issue of self-correlation may also be addressed in future, possibly, by devising optimal correlation functions to eliminate them, much as done in Ref.~\cite{De:2020yyx}. %That said, we emphasize that our analytical framework represents only one of many possible methods to compute two-particle correlation functions. One could simply couple the QGP evolution with a particle sampler that accounts for conservation laws, and evaluate $C_B$ or $C_Q$ in an event-by-event fashion. Similarly, one could use existing schemes evolving hydrodynamic fluctuations (with conserved currents) dynamically from noisy equations. 

We deem that calculations of angular and longitudinal correlations of net-charges will represent remarkable results in themselves, whose comparison with experimental data will open a new window onto the nontrivial thermal fluctuation properties of the quark-gluon plasma.

\section{Acknowledgments}

We thank the members of the C06 project \textit{``Flow and fluctuations in relativistic heavy-ion collisions''} within the ISOQUANT Collaborative Research Centre for useful discussion related to the topic of this work. We acknowledge useful discussions with Eduardo Grossi.  This work is supported by the Deutsche Forschungsgemeinschaft (DFG, German Research Foundation) under Germany's Excellence Strategy EXC 2181/1 - 390900948 (the Heidelberg STRUCTURES Excellence Cluster) and under 273811115 – SFB 1225 ISOQUANT as well as FL 736/3-1. This work is partially financed by the Baden-Württemberg Stiftung gGmbH.
\appendix

\section{Thermal correlation functions of an ideal gas}

\label{App:A}

We exhibit in this appendix the full derivation of the correlator $\langle \chi_n(\alpha) \chi_m(\alpha') \rangle_c$ in Eq.~(\ref{eq:finalChi}), i.e.,
\begin{equation}
\langle \chi_s(\alpha) \chi_t(\alpha^\prime) \rangle_c = \frac{1}{\sqrt{h(\alpha)}} \delta^{(3)}(\alpha-\alpha^\prime) \sigma_{st}(\alpha),
\label{eq:twoPointCorrelationVarphi}
\end{equation}
where we derive as well the thermal correlation matrix, $\sigma_{st}$.

\subsection{\label{Generating functionals for thermal fluctuations}Generating functionals for statistical fluctuations on the freeze-out surface}
Let us consider a three-dimensional hypersurface $\Sigma$ of four-dimensional space-time. We want to characterize statistical (e.\ g.\ thermal) fluctuations thereon. We assume that the hypersurface is parametrized by a coordinate system $\alpha^j$ with $j=1,2,3$. The embedding into space-time is given by the map $x^\mu(\alpha)$. The induced metric in the three-dimensional space is
\begin{equation}
h_{jk}(\alpha) = g_{\mu\nu} \left[ \frac{\partial}{\partial \alpha^j} x^\mu(\alpha) \right] \left[  \frac{\partial}{\partial \alpha^k} x^\nu(\alpha) \right],
\end{equation}
where $g_{\mu\nu}={\rm diag}(-,+,+,+)_{\mu\nu}$. We consider now that, for a given field configuration $\chi$, the probability density of fluctuations on the freeze-out surface, $\Sigma$, can be written as a functional
\begin{equation}
p[\chi] = \frac{1}{Z_\Sigma} e^{-I_\Sigma[\chi]},
\label{eq:DefProbDistrFunctional}
\end{equation}
with partition function
\begin{equation}
Z_\Sigma = \int D\chi \; e^{-I_\Sigma[\chi]}.
\end{equation}
In the last equation we have introduced the functional integral $\int D\chi = \int D \chi_1 \cdots \int D \chi_n$ on the surface $\Sigma$. The partition function on the freeze-out surface can be generalized to include external sources, $J_n$, 
\begin{equation}
Z_\Sigma[J] = \int D \chi \; \exp\left[ - I_\Sigma[\chi] + \int d^3 \alpha \sqrt{h} J_n(\alpha) \chi(\alpha) \right],
\label{eq:partitionFunctionSources}
\end{equation}
with $\sqrt{h} = \sqrt{\text{det}(h_{jk})}$ and the invariant hypersurface element $d^3\alpha \sqrt{h}$. Correlation functions are then obtained as functional derivatives of the partition function, e.\ g.\
\begin{equation}
\begin{split}
& \langle \chi_n(\alpha) \chi_m(\beta) \rangle = \\
& \frac{1}{Z_\Sigma [J]} \left( \frac{1}{\sqrt{h(\alpha)}} \frac{\delta}{\delta J_n(\alpha)} \right) \left( \frac{1}{\sqrt{h(\beta)}} \frac{\delta}{\delta J_m(\beta)} \right) Z_\Sigma [J]{\big |}_{J=0}.
\end{split}
\end{equation}
One may also introduce the generating functional for the connected correlation functions $W_\Sigma[J] = \ln Z_\Sigma[J]$ such that e.\ g.\ 
\begin{equation}
\begin{split}
& \langle \chi_n(\alpha) \chi_m(\beta) \rangle_c = \langle \chi_n(\alpha) \chi_m(\beta) \rangle - \langle \chi_n(\alpha) \rangle \langle \chi_m(\beta) \rangle \\
& = \left( \frac{1}{\sqrt{h(\alpha)}} \frac{\delta}{\delta J_n(\alpha)} \right) \left( \frac{1}{\sqrt{h(\beta)}} \frac{\delta}{\delta J_m(\beta)} \right) W[J]{\big |}_{J=0}.
\end{split}
\label{eq:ConnectedcorrelationfromW}
\end{equation}

All the relevant information about connected correlations on the freeze-out surface is encoded in $W_\Sigma[J]$. Determining this functional is of course a very nontrivial problem, already in, or close-to, thermal equilibrium for a strongly interacting theory like QCD. For vanishing baryon number chemical potential, one can in principle hope to obtain $W_\Sigma[J]$ from lattice QCD calculations. However, the determination of such local correlation functions would require prohibitive ``numerical statistics'' compared to the determination of global correlations, such as for integrated particle numbers or charges in some volume.

\subsection{\label{Local equilibrium approximation to correlation functions} Local equilibrium approximation to correlation functions}

To make progress, we assume thermal equilibrium in small fluid cells. The probability for $\chi$ is then determined by a change in entropy \cite{Landau:1980mil} \footnote{This may also be written in terms of a relative entropy \cite{Floerchinger:2020ogh}.}
\begin{equation}
    p[\chi]=\text{const}\times e^{\Delta S[\chi]},
    \label{StatphysProb}
\end{equation}
which can be directly related to eq.\ \eqref{eq:DefProbDistrFunctional},
\begin{equation}
I_\Sigma[\chi] = - \Delta S[\chi] + \text{const}~.
\label{eq:ActionDeltaEntropy}
\end{equation}
The additive constant is in fact irrelevant because it drops out in a calculation of correlation functions. For fluctuations of conserved quantities such as the four-momentum, $P^\mu$, and particle number, $N_j$, one can write locally
\begin{equation}
dS = \beta_\nu dP^\nu + \sum_j \alpha_j d N_j,
\end{equation}
with $\beta_\nu = u_\nu / T$, $T$ being the temperature, and $\alpha_j = \mu_j / T$ is the ratio of chemical potential, $\mu_j$, conjugate to the particle number $N_j$ and temperature $T$.

We split now the system into two parts - a subsystem labeled by an index $1$, and a bath labeled by $0$. We consider, then, that the change in entropy splits additively, $dS = dS_0 + dS_1$, such that
\begin{equation}
\begin{split}
dS = & dS_0 + dS_1 \\
= & \beta_{0,\nu} d P_0^\nu  + \beta_{1,\nu} d P_1^\nu + \sum_j \alpha_{j,0} d N_{j,0} +\sum_j \alpha_{j,1} d N_{j,1}, \\
= & \Delta\beta_\nu d P^\nu + \sum_j \Delta\alpha_j d N_j
\end{split}
\label{eq:differentialEntropy}
\end{equation}
where, to obtain the last line, we have used the conservation laws $dP_0^\nu+dP_1^\nu = 0$ and $dN_{j,0} + dN_{j,1}=0$, and set $\Delta\beta_\nu = \beta_{\nu,1} - \beta_{\nu,0}$ and $\Delta\alpha_j = \alpha_{j,1}-\alpha_{j,0}$. We also abbreviate $dP^\nu = d P^\nu_1$ and $dN_j = dN_{j,1}$. Now, we denote by $\Delta P^\nu$ and $\Delta N_j$ the amount of conserved quantities that are (temporarily) transmitted from the bath to the subsystem, while $\Delta\beta_\nu$ and $\Delta\alpha_j$ denote associated differences in conjugate fields. The latter are not independent of the former, but one can assume that $\Delta\beta_\nu$ and $\Delta\alpha_j$ are linear in $\Delta P^\nu$ and $\Delta N_j$, to lowest order in the variations. With this consideration, integration of Eq.\ \eqref{eq:differentialEntropy} leads to
\begin{equation}
\Delta S = \frac{1}{2} \left( \Delta\beta_\nu \Delta P^\nu + \sum_j \Delta \alpha_j \Delta N_j \right).
\label{eq:ChangeOfEntropyDecomp}
\end{equation}

In the next step, we assume that the above considerations hold for every small volume $d^3\alpha\sqrt{h}$ of the hypersurface $\Sigma$. This is a strong form of {\it local equilibrium hypothesis}, assuming that different volume elements are independent and that all correlation functions are short range. Keeping this in mind, from Eq.~(\ref{eq:ActionDeltaEntropy}) we obtain
\begin{equation}
I_\Sigma = -\Delta S = \frac{1}{2} \int d\Sigma_\mu \left\{ \Delta\beta_\nu \Delta T^{\mu\nu} + \sum_j \Delta\alpha_j \Delta N_j^\mu \right\},
\label{eq:functionalISigma}
\end{equation}
where we have employed the surface element $d\Sigma_\mu = d^3\alpha\sqrt{h} n_\mu$ with future pointing normal vector $n_\mu$ and we have used that locally $\Delta P^\nu = - \int d\Sigma_\mu \Delta T^{\mu\nu}$ with energy-momentum tensor $T^{\mu\nu}$ and similarly $\Delta N_j = - \int d\Sigma_\mu \Delta N_j^\mu$ with number current $N_j^\mu$ (the minus signs are again due to the metric signature $(-,+,+,+)$). 

The variation of the energy-momentum tensor, $\Delta T^{\mu\nu}$, and of the particle number current, $\Delta N^\mu$, can now be derived from their expression in the ideal fluid case considered throughout this manuscript
\begin{equation}
\label{eq:idefluid}
T^{\mu\nu} = \epsilon u^\mu u^\nu + p \Delta^{\mu\nu}, \quad\quad\quad N_j^\mu = n_j u^\mu.
\end{equation}
We choose as independent fields the three independent components of fluid velocity, $u^\mu$, the temperature, $T$, and the chemical potentials, $\mu_j$, and we consider that to linear order in $\Delta u^\mu$ one has $u_\mu\Delta u^\mu=0$. The linear variation in the energy momentum tensor is, consequently, given by
\begin{equation}
\begin{split}
&    \Delta T^{\mu \nu} = \left[ \left( T\tfrac{\partial ^2 p}{\partial T^2} +\sum \limits_{i} \mu_{i} \tfrac{\partial ^2 p}{\partial T \partial \mu_{i}} \right)u^{\mu} u^{\nu} +\tfrac{\partial p}{\partial T} \Delta^{\mu \nu} \right] \Delta T \\ 
    & +\sum\limits_{i} \left[\left( \sum_{j} \mu_{j} \tfrac{\partial ^2 p}{\partial \mu_{i} \partial \mu_{j}} + T \tfrac{\partial ^2 p}{\partial T \partial \mu_{i}} \right)u^{\mu} u^{\nu} +\tfrac{\partial p}{\partial \mu_{i}} \Delta^{\mu \nu} \right] \Delta \mu_{i}  \\
   & + \left[ \left(T \tfrac{\partial p}{\partial T} +\sum_{i} \mu_{i} \tfrac{\partial p}{\partial \mu_{i}} \right)\left(\delta_{\rho}^{\mu}u^{\nu}+u^{\mu}\delta_{\rho}^{\nu} \right) \right] \Delta u^{\rho},
\end{split}\label{eq:DeltaTmunuVariation1}
\end{equation}
while for the conserved particle currents
\begin{equation}
\begin{split}
    \Delta N^\mu_{i} = &  \left[\tfrac{\partial ^2 p}{\partial T \partial \mu_{i}} u^{\mu} \right]\Delta T + \sum \limits_{j} \left[\tfrac{\partial ^2 p}{ \partial \mu_{i} \partial \mu_{j}} u^{\mu} \right]\Delta \mu_{j} \\
    & + \left[ \tfrac{\partial  p}{\partial \mu_{i}} \delta_{\rho}^{\mu}\right] \Delta u^{\rho}    .
\end{split}\label{eq:DeltaNmuVariation1}
\end{equation}
Inserting these expression in Eq.~(\ref{eq:functionalISigma}) leads to a quadratic functional of the fluctuation fields
\begin{equation}
\begin{split}
   I_\Sigma  =  \frac{1}{2}\int d\Sigma_{\mu} {\Bigg \{} & \frac{u^{\mu}}{T} {\bigg (} \tfrac{\partial ^2 p}{\partial T^2} \Delta T^{2}  +2\sum\limits_{j} \tfrac{\partial ^2 p}{\partial T \partial \mu_{j}} \Delta T \Delta \mu_{j} \\
        & + \sum \limits_{i,j} \tfrac{\partial ^2 p}{ \partial \mu_{i} \partial \mu_{j}}  \Delta \mu_{i} \Delta \mu_{j} {\bigg )} \\
        & + 2 \frac{\Delta u^{\mu}}{T} {\bigg (} \tfrac{\partial p}{\partial T}\Delta T +\sum_{j} \tfrac{\partial p}{\partial \mu_{j}} \Delta \mu_{j} {\bigg)} \\ 
        & +\frac{u^{\mu}}{T} \left(\epsilon + p \right) \Delta_{\rho \sigma}\Delta u^{\rho} \Delta u^{\sigma} {\Bigg \}}.
\end{split}
\label{eq:Delta_S_fluct2}
\end{equation} 
We note that Eq.~\eqref{eq:Delta_S_fluct2}, when used in Eq.\ \eqref{eq:DefProbDistrFunctional}, defines a Gaussian probability distribution for the fields $\Delta T$, $\Delta\mu_j$ and the three independent components of $\Delta u^\mu$. We also note that the probability functional is ultra-local in the sense that it contains no spatial derivatives of the thermodynamic fields. To relax this assumption, one would have to go beyond the strong local equilibrium assumption, resorting for example to a derivative expansion.

\subsection{Thermal correlation matrix}

Being Gaussian, the functional integral defining the partition function in Eq.\ \eqref{eq:partitionFunctionSources} can easily be performed. Up to an irrelevant constant, one finds
\begin{equation}
Z_\Sigma[J] = \exp\left[ \frac{1}{2} \int d^3\alpha\sqrt{h} J_n(\alpha) \sigma_{nm}(\alpha) J_m(\alpha) \right],
\end{equation}
where $\sigma_{nm}(\alpha)$ is a matrix of local variances such that
\begin{equation}
\langle \chi_n(\alpha) \chi_m(\alpha^\prime) \rangle_c = \frac{1}{\sqrt{h(\alpha)}} \delta^{(3)}(\alpha-\alpha^\prime) \sigma_{nm}(\alpha).
\label{eq:twoPointCorrelationVarphi}
\end{equation}
One can in fact directly read of the inverse of $\sigma_{nm}$ from equation \eqref{eq:Delta_S_fluct2} by writing
\begin{equation}
I_\Sigma[\chi] = \frac{1}{2} \int d^3\alpha \sqrt{h} \, \chi_n(\alpha) (\sigma^{-1})_{nm} \chi_m(\alpha).
\label{eq:ISigmaFunctional}
\end{equation}
Choosing as variables $\chi_n=(\Delta T, \Delta\mu_i, \Delta u^\rho)$ and taking into account that $d\Sigma_\mu=d^3 \alpha\sqrt{h} \, n_\mu$ with normal vector $n_\mu$ leads to 
\begin{equation}
\sigma^{-1} = \begin{pmatrix} - \frac{n\cdot u}{T} \frac{\partial^2 p}{\partial T^2} & -\frac{n\cdot u}{T} \frac{\partial^2 p}{\partial T \partial \mu_j} & - \frac{n_\sigma}{T}\frac{\partial p}{\partial T} \\ 
- \frac{n\cdot u}{T} \frac{\partial^2 p}{\partial T \partial\mu_i} & - \frac{n \cdot u}{T} \frac{\partial^2 p}{\partial\mu_i \partial \mu_j} & - \frac{n_\sigma}{T} \frac{\partial p}{\partial \mu_j} \\ - \frac{n_\rho}{T} \frac{\partial p}{\partial T} & - \frac{n_\rho}{T} \frac{\partial p}{\partial \mu_i} & - \frac{n\cdot u}{T}(\epsilon+p) \Delta_{\rho\sigma}
\label{eq:ThermalCorrelationMatrixInverted}
\end{pmatrix}.
\end{equation}
This matrix can easily be inverted in a concrete application. Note that its entries are determined by the equation of state of the considered system.


\begin{thebibliography}{99}
%\cite{Braun-Munzinger:2007edi}
\bibitem{Braun-Munzinger:2007edi}
P.~Braun-Munzinger and J.~Stachel,
%``The quest for the quark-gluon plasma,''
Nature \textbf{448}, 302-309 (2007)
doi:10.1038/nature06080
%194 citations counted in INSPIRE as of 10 Dec 2021

%\cite{Busza:2018rrf}
\bibitem{Busza:2018rrf}
W.~Busza, K.~Rajagopal and W.~van der Schee,
%``Heavy Ion Collisions: The Big Picture, and the Big Questions,''
Ann. Rev. Nucl. Part. Sci. \textbf{68}, 339-376 (2018)
doi:10.1146/annurev-nucl-101917-020852
[arXiv:1802.04801 [hep-ph]].
%298 citations counted in INSPIRE as of 17 Dec 2021

%\cite{Romatschke:2017ejr}
\bibitem{Romatschke:2017ejr}
P.~Romatschke and U.~Romatschke,
%``Relativistic Fluid Dynamics In and Out of Equilibrium,''
doi:10.1017/9781108651998
[arXiv:1712.05815 [nucl-th]].
%202 citations counted in INSPIRE as of 14 Dec 2021

%\cite{Schlichting:2019abc}
\bibitem{Schlichting:2019abc}
S.~Schlichting and D.~Teaney,
%``The First fm/c of Heavy-Ion Collisions,''
Ann. Rev. Nucl. Part. Sci. \textbf{69}, 447-476 (2019)
doi:10.1146/annurev-nucl-101918-023825
[arXiv:1908.02113 [nucl-th]].
%40 citations counted in INSPIRE as of 03 Dec 2021

%\cite{Berges:2020fwq}
\bibitem{Berges:2020fwq}
J.~Berges, M.~P.~Heller, A.~Mazeliauskas and R.~Venugopalan,
%``QCD thermalization: Ab initio approaches and interdisciplinary connections,''
Rev. Mod. Phys. \textbf{93}, no.3, 035003 (2021)
doi:10.1103/RevModPhys.93.035003
[arXiv:2005.12299 [hep-th]].
%69 citations counted in INSPIRE as of 03 Dec 2021

%\cite{Luzum:2013yya}
\bibitem{Luzum:2013yya}
M.~Luzum and H.~Petersen,
%``Initial State Fluctuations and Final State Correlations in Relativistic Heavy-Ion Collisions,''
J. Phys. G \textbf{41}, 063102 (2014)
doi:10.1088/0954-3899/41/6/063102
[arXiv:1312.5503 [nucl-th]].
%137 citations counted in INSPIRE as of 14 Dec 2021

%\cite{Cheng:2008zh}
\bibitem{Cheng:2008zh}
M.~Cheng, P.~Hegde, C.~Jung, F.~Karsch, O.~Kaczmarek, E.~Laermann, R.~D.~Mawhinney, C.~Miao, P.~Petreczky and C.~Schmidt, \textit{et al.}
%``Baryon Number, Strangeness and Electric Charge Fluctuations in QCD at High Temperature,''
Phys. Rev. D \textbf{79}, 074505 (2009)
doi:10.1103/PhysRevD.79.074505
[arXiv:0811.1006 [hep-lat]].
%309 citations counted in INSPIRE as of 03 Dec 2021

%\cite{Fukushima:2010bq}
\bibitem{Fukushima:2010bq}
K.~Fukushima and T.~Hatsuda,
%``The phase diagram of dense QCD,''
Rept. Prog. Phys. \textbf{74}, 014001 (2011)
doi:10.1088/0034-4885/74/1/014001
[arXiv:1005.4814 [hep-ph]].
%713 citations counted in INSPIRE as of 11 Dec 2021

%\cite{Bzdak:2019pkr}
\bibitem{Bzdak:2019pkr}
A.~Bzdak, S.~Esumi, V.~Koch, J.~Liao, M.~Stephanov and N.~Xu,
%``Mapping the Phases of Quantum Chromodynamics with Beam Energy Scan,''
Phys. Rept. \textbf{853}, 1-87 (2020)
doi:10.1016/j.physrep.2020.01.005
[arXiv:1906.00936 [nucl-th]].
%164 citations counted in INSPIRE as of 17 Dec 2021

%\cite{Bluhm:2020mpc}
\bibitem{Bluhm:2020mpc}
M.~Bluhm, A.~Kalweit, M.~Nahrgang, M.~Arslandok, P.~Braun-Munzinger, S.~Floerchinger, E.~S.~Fraga, M.~Gazdzicki, C.~Hartnack and C.~Herold, \textit{et al.}
%``Dynamics of critical fluctuations: Theory \textendash{} phenomenology \textendash{} heavy-ion collisions,''
Nucl. Phys. A \textbf{1003}, 122016 (2020)
doi:10.1016/j.nuclphysa.2020.122016
[arXiv:2001.08831 [nucl-th]].
%47 citations counted in INSPIRE as of 14 Dec 2021

%\cite{Kapusta:2011gt}
\bibitem{Kapusta:2011gt}
J.~I.~Kapusta, B.~Muller and M.~Stephanov,
%``Relativistic Theory of Hydrodynamic Fluctuations with Applications to Heavy Ion Collisions,''
Phys. Rev. C \textbf{85}, 054906 (2012)
doi:10.1103/PhysRevC.85.054906
[arXiv:1112.6405 [nucl-th]].
%150 citations counted in INSPIRE as of 10 Dec 2021

%\cite{Young:2013fka}
\bibitem{Young:2013fka}
C.~Young,
%``Numerical integration of thermal noise in relativistic hydrodynamics,''
Phys. Rev. C \textbf{89}, no.2, 024913 (2014)
doi:10.1103/PhysRevC.89.024913
[arXiv:1306.0472 [nucl-th]].
%24 citations counted in INSPIRE as of 03 Dec 2021

%\cite{Ling:2013ksb}
\bibitem{Ling:2013ksb}
B.~Ling, T.~Springer and M.~Stephanov,
%``Hydrodynamics of charge fluctuations and balance functions,''
Phys. Rev. C \textbf{89}, no.6, 064901 (2014)
doi:10.1103/PhysRevC.89.064901
[arXiv:1310.6036 [nucl-th]].
%26 citations counted in INSPIRE as of 03 Dec 2021

%\cite{Young:2014pka}
\bibitem{Young:2014pka}
C.~Young, J.~I.~Kapusta, C.~Gale, S.~Jeon and B.~Schenke,
%``Thermally Fluctuating Second-Order Viscous Hydrodynamics and Heavy-Ion Collisions,''
Phys. Rev. C \textbf{91}, no.4, 044901 (2015)
doi:10.1103/PhysRevC.91.044901
[arXiv:1407.1077 [nucl-th]].
%47 citations counted in INSPIRE as of 03 Dec 2021

%\cite{Murase:2015oie}
\bibitem{Murase:2015oie}
K.~Murase,
%``Causal hydrodynamic fluctuations and their effects on high-energy nuclear collisions,''
doi:10.15083/00072981
%1 citations counted in INSPIRE as of 03 Dec 2021

%\cite{Yan:2015lfa}
\bibitem{Yan:2015lfa}
L.~Yan and H.~Gr\"onqvist,
%``Hydrodynamical noise and Gubser flow,''
JHEP \textbf{03}, 121 (2016)
doi:10.1007/JHEP03(2016)121
[arXiv:1511.07198 [nucl-th]].
%23 citations counted in INSPIRE as of 03 Dec 2021

%\cite{Nagai:2016wyx}
\bibitem{Nagai:2016wyx}
K.~Nagai, R.~Kurita, K.~Murase and T.~Hirano,
%``Causal hydrodynamic fluctuation in Bjorken expansion,''
Nucl. Phys. A \textbf{956}, 781-784 (2016)
doi:10.1016/j.nuclphysa.2016.02.007
[arXiv:1602.00794 [nucl-th]].
%13 citations counted in INSPIRE as of 03 Dec 2021

%\cite{Gavin:2016hmv}
\bibitem{Gavin:2016hmv}
S.~Gavin, G.~Moschelli and C.~Zin,
%``Rapidity Correlation Structure in Nuclear Collisions,''
Phys. Rev. C \textbf{94}, no.2, 024921 (2016)
doi:10.1103/PhysRevC.94.024921
[arXiv:1606.02692 [nucl-th]].
%18 citations counted in INSPIRE as of 05 Dec 2021

%\cite{Akamatsu:2016llw}
\bibitem{Akamatsu:2016llw}
Y.~Akamatsu, A.~Mazeliauskas and D.~Teaney,
%``A kinetic regime of hydrodynamic fluctuations and long time tails for a Bjorken expansion,''
Phys. Rev. C \textbf{95}, no.1, 014909 (2017)
doi:10.1103/PhysRevC.95.014909
[arXiv:1606.07742 [nucl-th]].
%68 citations counted in INSPIRE as of 03 Dec 2021

%\cite{Sakai:2017rfi}
\bibitem{Sakai:2017rfi}
A.~Sakai, K.~Murase and T.~Hirano,
%``Hydrodynamic fluctuations in Pb + Pb collisions at LHC,''
Nucl. Phys. A \textbf{967}, 445-448 (2017)
doi:10.1016/j.nuclphysa.2017.05.010
%7 citations counted in INSPIRE as of 03 Dec 2021

%\cite{Kapusta:2017hfi}
\bibitem{Kapusta:2017hfi}
J.~I.~Kapusta and C.~Plumberg,
%``Causal Electric Charge Diffusion and Balance Functions in Relativistic Heavy Ion Collisions,''
Phys. Rev. C \textbf{97}, no.1, 014906 (2018)
[erratum: Phys. Rev. C \textbf{102}, no.1, 019901 (2020)]
doi:10.1103/PhysRevC.97.014906
[arXiv:1710.03329 [nucl-th]].
%19 citations counted in INSPIRE as of 03 Dec 2021

%\cite{Chattopadhyay:2017rgh}
\bibitem{Chattopadhyay:2017rgh}
C.~Chattopadhyay, R.~S.~Bhalerao and S.~Pal,
%``Thermal noise in a boost-invariant matter expansion in relativistic heavy-ion collisions,''
Phys. Rev. C \textbf{97}, no.5, 054902 (2018)
doi:10.1103/PhysRevC.97.054902
[arXiv:1711.10759 [nucl-th]].
%4 citations counted in INSPIRE as of 03 Dec 2021

%\cite{Stephanov:2017ghc}
\bibitem{Stephanov:2017ghc}
M.~Stephanov and Y.~Yin,
%``Hydrodynamics with parametric slowing down and fluctuations near the critical point,''
Phys. Rev. D \textbf{98}, no.3, 036006 (2018)
doi:10.1103/PhysRevD.98.036006
[arXiv:1712.10305 [nucl-th]].
%87 citations counted in INSPIRE as of 17 Dec 2021

%\cite{Singh:2018dpk}
\bibitem{Singh:2018dpk}
M.~Singh, C.~Shen, S.~McDonald, S.~Jeon and C.~Gale,
%``Hydrodynamic Fluctuations in Relativistic Heavy-Ion Collisions,''
Nucl. Phys. A \textbf{982}, 319-322 (2019)
doi:10.1016/j.nuclphysa.2018.10.061
[arXiv:1807.05451 [nucl-th]].
%25 citations counted in INSPIRE as of 03 Dec 2021

%\cite{Akamatsu:2018vjr}
\bibitem{Akamatsu:2018vjr}
Y.~Akamatsu, D.~Teaney, F.~Yan and Y.~Yin,
%``Transits of the QCD critical point,''
Phys. Rev. C \textbf{100}, no.4, 044901 (2019)
doi:10.1103/PhysRevC.100.044901
[arXiv:1811.05081 [nucl-th]].
%33 citations counted in INSPIRE as of 17 Dec 2021

%\cite{An:2019osr}
\bibitem{An:2019osr}
X.~An, G.~Basar, M.~Stephanov and H.~U.~Yee,
%``Relativistic Hydrodynamic Fluctuations,''
Phys. Rev. C \textbf{100}, no.2, 024910 (2019)
doi:10.1103/PhysRevC.100.024910
[arXiv:1902.09517 [hep-th]].
%34 citations counted in INSPIRE as of 03 Dec 2021

%\cite{Murase:2019cwc}
\bibitem{Murase:2019cwc}
K.~Murase,
%``Causal hydrodynamic fluctuations in non-static and inhomogeneous backgrounds,''
Annals Phys. \textbf{411}, 167969 (2019)
doi:10.1016/j.aop.2019.167969
[arXiv:1904.11217 [nucl-th]].
%5 citations counted in INSPIRE as of 03 Dec 2021

%\cite{An:2019csj}
\bibitem{An:2019csj}
X.~An, G.~Ba\c{s}ar, M.~Stephanov and H.~U.~Yee,
%``Fluctuation dynamics in a relativistic fluid with a critical point,''
Phys. Rev. C \textbf{102}, no.3, 034901 (2020)
doi:10.1103/PhysRevC.102.034901
[arXiv:1912.13456 [hep-th]].
%20 citations counted in INSPIRE as of 03 Dec 2021

%\cite{De:2020yyx}
\bibitem{De:2020yyx}
A.~De, C.~Plumberg and J.~I.~Kapusta,
%``Calculating Fluctuations and Self-Correlations Numerically for Causal Charge Diffusion in Relativistic Heavy-Ion Collisions,''
Phys. Rev. C \textbf{102}, no.2, 024905 (2020)
doi:10.1103/PhysRevC.102.024905
[arXiv:2003.04878 [nucl-th]].
%3 citations counted in INSPIRE as of 10 Dec 2021

%\cite{Sakai:2020pjw}
\bibitem{Sakai:2020pjw}
A.~Sakai, K.~Murase and T.~Hirano,
%``Rapidity decorrelation of anisotropic flow caused by hydrodynamic fluctuations,''
Phys. Rev. C \textbf{102}, no.6, 064903 (2020)
doi:10.1103/PhysRevC.102.064903
[arXiv:2003.13496 [nucl-th]].
%11 citations counted in INSPIRE as of 03 Dec 2021

%\cite{Chao:2020kcf}
\bibitem{Chao:2020kcf}
J.~Chao and T.~Schaefer,
%``Multiplicative noise and the diffusion of conserved densities,''
JHEP \textbf{01}, 071 (2021)
doi:10.1007/JHEP01(2021)071
[arXiv:2008.01269 [hep-th]].
%5 citations counted in INSPIRE as of 10 Dec 2021

%\cite{Sakai:2021pev}
\bibitem{Sakai:2021pev}
A.~Sakai, K.~Murase and T.~Hirano,
%``Effects of hydrodynamic and initial longitudinal fluctuations on rapidity decorrelation of collective flow,''
[arXiv:2111.08963 [nucl-th]].
%0 citations counted in INSPIRE as of 03 Dec 2021

%\cite{Cooper:1974mv}
\bibitem{Cooper:1974mv}
F.~Cooper and G.~Frye,
%``Comment on the Single Particle Distribution in the Hydrodynamic and Statistical Thermodynamic Models of Multiparticle Production,''
Phys. Rev. D \textbf{10}, 186 (1974)
doi:10.1103/PhysRevD.10.186
%1000 citations counted in INSPIRE as of 17 Dec 2021

%\cite{Teaney:2003kp}
\bibitem{Teaney:2003kp}
D.~Teaney,
%``The Effects of viscosity on spectra, elliptic flow, and HBT radii,''
Phys. Rev. C \textbf{68}, 034913 (2003)
doi:10.1103/PhysRevC.68.034913
[arXiv:nucl-th/0301099 [nucl-th]].
%760 citations counted in INSPIRE as of 03 Dec 2021

%\cite{Paquet:2015lta}
\bibitem{Paquet:2015lta}
J.~F.~Paquet, C.~Shen, G.~S.~Denicol, M.~Luzum, B.~Schenke, S.~Jeon and C.~Gale,
%``Production of photons in relativistic heavy-ion collisions,''
Phys. Rev. C \textbf{93}, no.4, 044906 (2016)
doi:10.1103/PhysRevC.93.044906
[arXiv:1509.06738 [hep-ph]].
%197 citations counted in INSPIRE as of 05 Dec 2021

%\cite{Calzetta:2008iqa}
\bibitem{Calzetta:2008iqa}
E.~A.~Calzetta and B.~L.~B.~Hu,
%``Nonequilibrium Quantum Field Theory,''
doi:10.1017/CBO9780511535123
%37 citations counted in INSPIRE as of 03 Dec 2021

%\cite{Bjorken:1982qr}
\bibitem{Bjorken:1982qr}
J.~D.~Bjorken,
%``Highly Relativistic Nucleus-Nucleus Collisions: The Central Rapidity Region,''
Phys. Rev. D \textbf{27}, 140-151 (1983)
doi:10.1103/PhysRevD.27.140
%3318 citations counted in INSPIRE as of 17 Dec 2021

%\cite{Luzum:2011mm}
\bibitem{Luzum:2011mm}
M.~Luzum,
%``Flow fluctuations and long-range correlations: elliptic flow and beyond,''
J. Phys. G \textbf{38}, 124026 (2011)
doi:10.1088/0954-3899/38/12/124026
[arXiv:1107.0592 [nucl-th]].
%74 citations counted in INSPIRE as of 03 Dec 2021

%\cite{Stephanov:1998dy}
\bibitem{Stephanov:1998dy}
M.~A.~Stephanov, K.~Rajagopal and E.~V.~Shuryak,
%``Signatures of the tricritical point in QCD,''
Phys. Rev. Lett. \textbf{81}, 4816-4819 (1998)
doi:10.1103/PhysRevLett.81.4816
[arXiv:hep-ph/9806219 [hep-ph]].
%1087 citations counted in INSPIRE as of 17 Dec 2021

%\cite{Schwarz:2017bdg}
\bibitem{Schwarz:2017bdg}
C.~Schwarz, D.~Oliinychenko, L.~G.~Pang, S.~Ryu and H.~Petersen,
%``Different realizations of Cooper\textendash{}Frye sampling with conservation laws,''
J. Phys. G \textbf{45}, no.1, 015001 (2018)
doi:10.1088/1361-6471/aa90eb
[arXiv:1707.07026 [hep-ph]].
%11 citations counted in INSPIRE as of 03 Dec 2021

%\cite{Oliinychenko:2019zfk}
\bibitem{Oliinychenko:2019zfk}
D.~Oliinychenko and V.~Koch,
%``Microcanonical Particlization with Local Conservation Laws,''
Phys. Rev. Lett. \textbf{123}, no.18, 182302 (2019)
doi:10.1103/PhysRevLett.123.182302
[arXiv:1902.09775 [hep-ph]].
%28 citations counted in INSPIRE as of 17 Dec 2021

%\cite{Oliinychenko:2020cmr}
\bibitem{Oliinychenko:2020cmr}
D.~Oliinychenko, S.~Shi and V.~Koch,
%``Effects of local event-by-event conservation laws in ultrarelativistic heavy-ion collisions at particlization,''
Phys. Rev. C \textbf{102}, no.3, 034904 (2020)
doi:10.1103/PhysRevC.102.034904
[arXiv:2001.08176 [hep-ph]].
%14 citations counted in INSPIRE as of 17 Dec 2021

%\cite{Vovchenko:2021yen}
\bibitem{Vovchenko:2021yen}
V.~Vovchenko,
%``Correcting event-by-event fluctuations in heavy-ion collisions for exact global conservation laws with the generalized subensemble acceptance method,''
[arXiv:2106.13775 [hep-ph]].
%4 citations counted in INSPIRE as of 03 Dec 2021

%\cite{Schnedermann:1993ws}
\bibitem{Schnedermann:1993ws}
E.~Schnedermann, J.~Sollfrank and U.~W.~Heinz,
%``Thermal phenomenology of hadrons from 200-A/GeV S+S collisions,''
Phys. Rev. C \textbf{48}, 2462-2475 (1993)
doi:10.1103/PhysRevC.48.2462
[arXiv:nucl-th/9307020 [nucl-th]].
%1028 citations counted in INSPIRE as of 16 Dec 2021

%\cite{Vovchenko:2014pka}
\bibitem{Vovchenko:2014pka}
V.~Vovchenko, D.~V.~Anchishkin and M.~I.~Gorenstein,
%``Hadron Resonance Gas Equation of State from Lattice QCD,''
Phys. Rev. C \textbf{91}, no.2, 024905 (2015)
doi:10.1103/PhysRevC.91.024905
[arXiv:1412.5478 [nucl-th]].
%58 citations counted in INSPIRE as of 09 Dec 2021

%\cite{Mazeliauskas:2019ifr}
\bibitem{Mazeliauskas:2019ifr}
A.~Mazeliauskas and V.~Vislavicius,
%``Temperature and fluid velocity on the freeze-out surface from $\pi$, $K$, $p$ spectra in pp, p-Pb and Pb-Pb collisions,''
Phys. Rev. C \textbf{101}, no.1, 014910 (2020)
doi:10.1103/PhysRevC.101.014910
[arXiv:1907.11059 [hep-ph]].
%18 citations counted in INSPIRE as of 03 Dec 2021

%\cite{ParticleDataGroup:2020ssz}
\bibitem{ParticleDataGroup:2020ssz}
P.~A.~Zyla \textit{et al.} [Particle Data Group],
%``Review of Particle Physics,''
PTEP \textbf{2020}, no.8, 083C01 (2020)
doi:10.1093/ptep/ptaa104
%2493 citations counted in INSPIRE as of 17 Dec 2021

%\cite{Pratt:2012dz}
\bibitem{Pratt:2012dz}
S.~Pratt,
%``Identifying the Charge Carriers of the Quark-Gluon Plasma,''
Phys. Rev. Lett. \textbf{108}, 212301 (2012)
doi:10.1103/PhysRevLett.108.212301
[arXiv:1203.4578 [nucl-th]].
%36 citations counted in INSPIRE as of 03 Dec 2021

%\cite{Kitazawa:2017ljq}
\bibitem{Kitazawa:2017ljq}
M.~Kitazawa and X.~Luo,
%``Properties and uses of factorial cumulants in relativistic heavy-ion collisions,''
Phys. Rev. C \textbf{96}, no.2, 024910 (2017)
doi:10.1103/PhysRevC.96.024910
[arXiv:1704.04909 [nucl-th]].
%24 citations counted in INSPIRE as of 03 Dec 2021

%\cite{Landau:1980mil}
\bibitem{Landau:1980mil}
L.~D.~Landau and E.~M.~Lifshitz,
\textit{``Statistical Physics, Part 1,''}
ISBN: 9780750633727,
Butterworth-Heinemann, Oxford, 1980 
%14 citations counted in INSPIRE as of 03 Dec 2021

%\cite{Floerchinger:2020ogh}
\bibitem{Floerchinger:2020ogh}
S.~Floerchinger and T.~Haas,
%``Thermodynamics from relative entropy,''
Phys. Rev. E \textbf{102}, no.5, 052117 (2020)
doi:10.1103/PhysRevE.102.052117
[arXiv:2004.13533 [cond-mat.stat-mech]].
%3 citations counted in INSPIRE as of 03 Dec 2021

\end{thebibliography}
\end{document}